%% file: manuscript.tex
\documentclass[aip,jcp,amsmath,amssymb,floatfix,reprint,citeautoscript,articletitle=false,noeprint]{revtex4-1}
\usepackage{siunitx}
\DeclareSIUnit\calorie{cal}
\usepackage{graphicx}
\usepackage{wrapfig}
\usepackage{xcolor}
\usepackage{caption}
\usepackage{subcaption}
\usepackage[colorlinks,allcolors=black,citecolor=blue,urlcolor=blue]{hyperref}

\graphicspath{{figures/}}

\begin{document}

\title
{%
Force-induced Catastrophes on Energy Landscapes:
Mechanochemical Manipulation of 
Downhill and Uphill Bifurcations 
Explains Ring-opening Selectivity of Cyclopropanes
}%
\date{\today}

\author{Miriam Wollenhaupt}
\affiliation{%
Lehrstuhl f\"{u}r Theoretische Chemie,
Ruhr--Universit\"{a}t Bochum, 44780 Bochum, Germany}
\author{Christoph Schran}
\affiliation{%
Lehrstuhl f\"{u}r Theoretische Chemie,
Ruhr--Universit\"{a}t Bochum, 44780 Bochum, Germany}
\author{Martin Krupi\v{c}ka}
\thanks{%
Present Address:
Department of Organic Chemistry, University of Chemistry and Technology,
Technicka~5, 16628~Prague, Czech Republic}
\affiliation{%
Lehrstuhl f\"{u}r Theoretische Chemie,
Ruhr--Universit\"{a}t Bochum, 44780 Bochum, Germany}
\author{Dominik Marx}
\email{dominik.marx@rub.de}
\affiliation{%
Lehrstuhl f\"{u}r Theoretische Chemie,
Ruhr--Universit\"{a}t Bochum, 44780 Bochum, Germany}


\begin{abstract}
\begin{wrapfigure}{r}{0.3\textwidth}
  \hspace{-1.5cm}
  \includegraphics[width=0.3\textwidth]{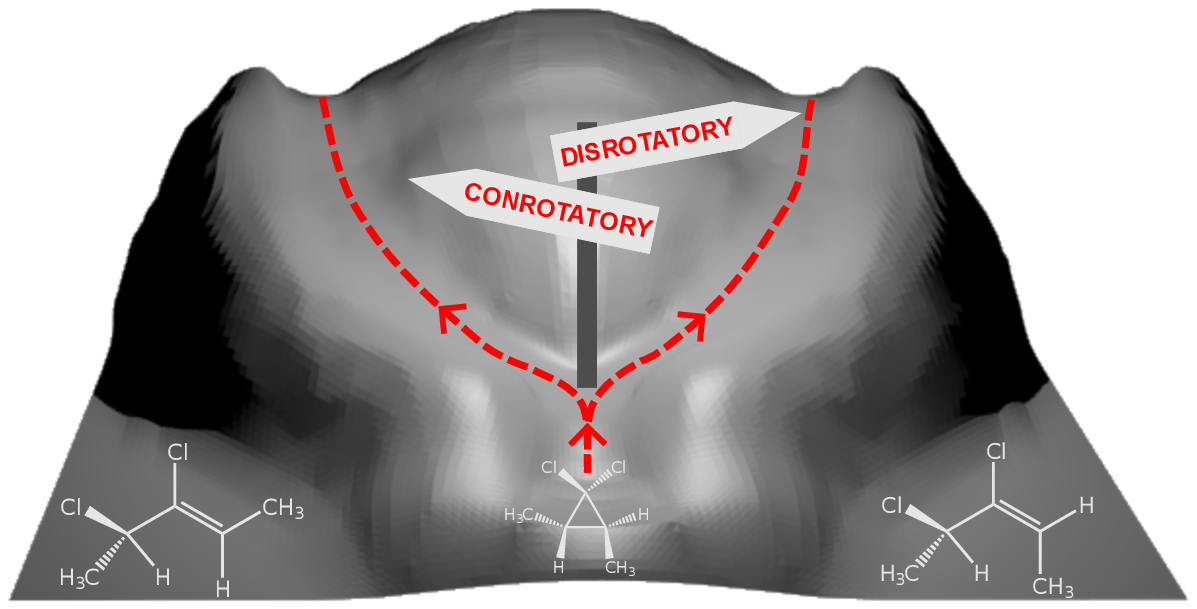}
\end{wrapfigure}
The mechanochemistry of ring-opening reactions of cyclopropane derivatives
turns out to be unexpectedly rich and puzzling.
After 
showing
that a rare so-called uphill bifurcation in the case of
\textit{trans}-\textit{gem}-difluorocyclopropane
turns into a downhill bifurcation upon substitution of fluorine by
chlorine, bromine and iodine in the thermal activation limit, the 
dichloro derivative is studied systematically
in the realm of mechanochemical activation.
Detailed exploration
of the force-transformed potential energy surface
of \textit{trans}-\textit{gem}-dichlorocyclopropane
in terms of Dijkstra path analysis 
unveils a hitherto unknown topological catastrophe
where the global shape of the energy landscape is 
fundamentally changed.
From 
thermal activation
up to moderately large forces, it is an uphill bifurcation that 
decides about
dis-
{\em versus} conrotatory ring-opening 
followed by separate transition states along both pathways.
Above 
a
critical force, the two distinct
transition states merge to yield a single transition state such that the
decision about the dis- {\em versus} conrotatory ring-opening process
is taken 
at a newly established downhill bifurcation.
The discovery of a force-induced qualitative change of the topology
of a reaction network
vastly transcends
the previous understanding of the ring-opening reaction of this species.
It would be astonishing to not discover a wealth of such 
catastrophes for mechanochemically activated reactions
which will greatly extend the known opportunities to
manipulate chemical reaction networks.
\end{abstract}
\maketitle
\section{Introduction}
\label{sec:Introduction}
Chemists usually think of a transition state as a 
stationary-point structure connecting a reactant with one 
product state, thus linking two minima on the potential energy surface (PES).
However, it is well established that 
PESs can also feature bifurcations which
open the opportunity to explore  pathways 
that start at the same reactant, 
share the same path for a while, but lead to more than one reaction channel
at some point, thus connecting the same reactant to more than one 
product state.\cite{Ess2008,Rehbein2011}
As will be described in more detail below, such 
bifurcations along a reaction pathway
usually occur after a transition state, leading to 
what is called ``downhill bifurcations'', but in rare cases so-called ``uphill bifurcations'' 
occur before even reaching a transition state.
Chemical substitution represents the classic way 
to change the PES and, thus, to manipulate also such bifurcations.
Within this traditional approach, 
the underlying energy landscape gets changed due to steric and
other electronic effects which can offer the possibility to arrive at new reaction 
pathways.\cite{Jones2001,Limanto2003,Wang2005,Katori2010}
Obviously, these manipulations can only be realized in a discrete manner by
exchanging one atom or a group of atoms by another one. 

In stark contrast, it is well established in
the framework of covalent mechanochemistry~\cite{Beyer2005,Caruso2009,Ribas-Arino2012,Stauch2016} 
that potential energy landscapes, i.e. PESs, can be 
{\em continuously} distorted as a result of applying external forces to 
molecules.\cite{Ong2009,Ribas-Arino2009,%
Ribas-Arino2012}
As reviewed in depth from the experimental~\cite{Caruso2009}
and computational~\cite{Ribas-Arino2012} viewpoints, 
mechanochemical activation
can lead to different products from the usual thermally 
or photochemically favored ones, 
if sufficiently large forces are applied.
In the following, we will explore in detail for a specific system class,
namely for cyclopropanes, how bifurcations can be systematically
tuned by means of applying such mechanical forces to molecules.

\begin{figure}
	\begin{center}
	\centering
	\begin{subfigure}{0.23\textwidth}
	\centering
		\includegraphics[width=\textwidth]{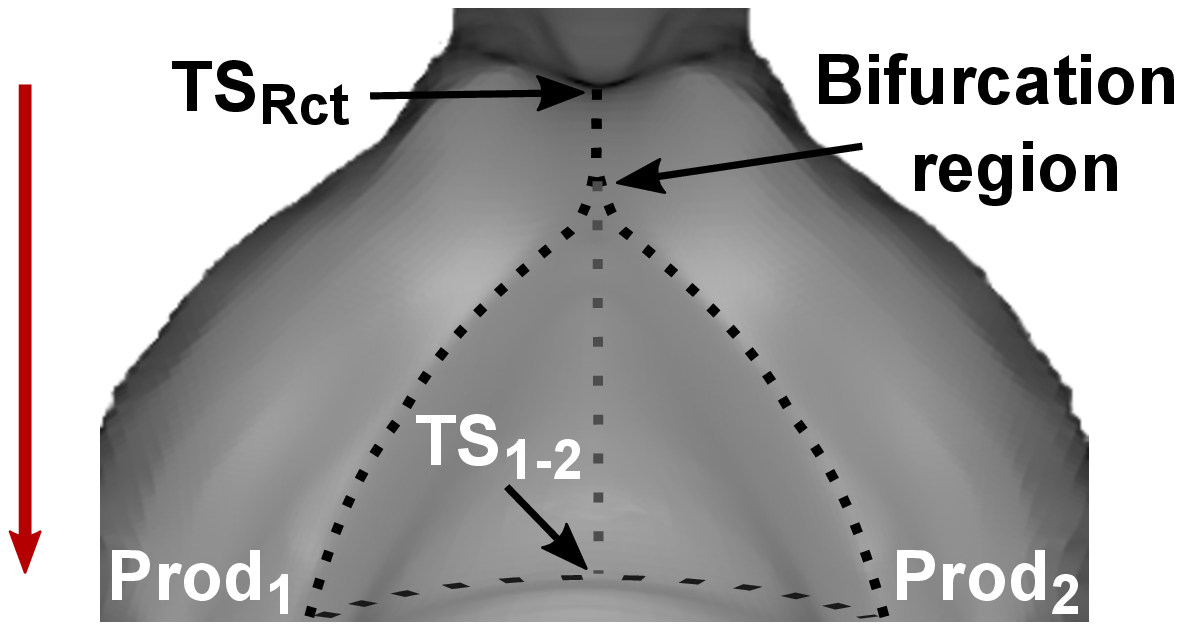}
		\subcaption{Downhill}
		\label{fig:Bifurcations_Generell_Downhill}
	\end{subfigure}
	\begin{subfigure}{0.23\textwidth}
	\centering
		\includegraphics[width=\textwidth]{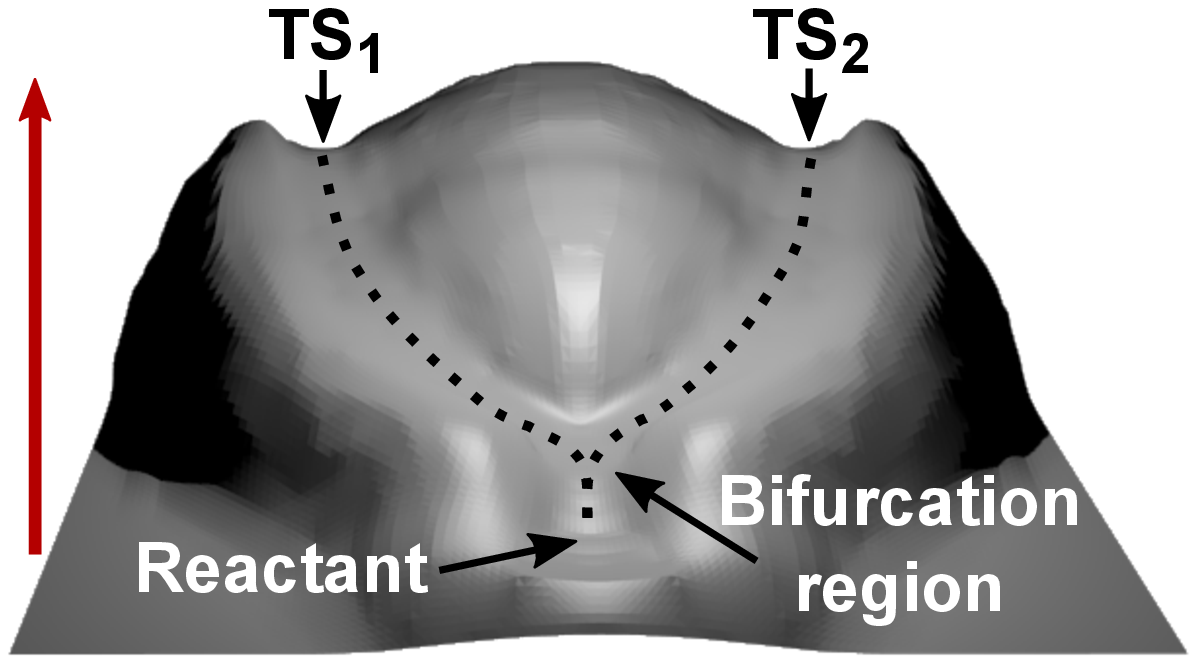}
		\subcaption{Uphill}
		\label{fig:Bifurcations_Generell_Uphill}
	\end{subfigure}
    	\caption{%
Schematic potential energy surfaces with 
downhill (a) and uphill (b) bifurcations. 
The reactions in (a) and (b) proceed from top to bottom
and from bottom to top, respectively, as indicated by the
two vertical arrows to the left of the landscapes. 
}
	\label{fig:Bifurcations_Generell}
	\end{center}
\end{figure}
A scenario that is rather often found in the literature are
bifurcations in the post-transition state region 
as sketched in Fig.~\ref{fig:Bifurcations_Generell}(a), 
which are usually denoted as downhill bifurcations.
The common pathway that starts in the reactant state ascends in energy
(not visible in the figure)
until it reaches the transition state $\textrm{TS}_{\textrm{Rct}}$ 
and from there on it exclusively moves downhill in energy 
(from top to bottom in Fig.~\ref{fig:Bifurcations_Generell}(a)) 
as in any standard chemical reaction.
At some distinct point upon moving toward the product state, however,
the common path
arrives at the bifurcation where it splits into two separate pathways
which ultimately lead to two different minima denoted as
Prod$_{1}$ and Prod$_{2}$
in Fig.~\ref{fig:Bifurcations_Generell}(a).
These two products, in turn, are connected by a second
transition state, $\textrm{TS}_{\textrm{1-2}}$,
which is related to the direct interconversion of Product~1 and~2.  
Thus, a PES exhibiting a downhill bifurcation possesses 
two close-lying and consecutive transition states with no 
local minimum in between.
Much more seldomly found, and thus of considerable fundamental interest, 
are uphill bifurcations such as the one sketched 
in Fig.~\ref{fig:Bifurcations_Generell}(b).
An uphill bifurcation is best described by a reaction path 
that starts at the reactant minimum and leads uphill in energy
while initially passing through a unique valley
akin to any standard chemical reaction
thereby moving from bottom to top in Fig.~\ref{fig:Bifurcations_Generell}(b).
At some specific point, however, the valley branches 
%
%
%
at the bifurcation
{\em before} any transition state is reached
%
%
and two new valleys with their own subsequent transition states,
labeled as TS$_1$ and TS$_2$ in Fig.~\ref{fig:Bifurcations_Generell}(b), 
emerge.
Although downhill bifurcations are, for instance, well-established in the realm of
thermal cycloaddition reactions~\cite{Yu2015,Yu2017},
uphill bifurcations are far less 
frequently described in the extant literature~\cite{Valtazanos1986,Baker1988,Quapp2001,Windhorn2003,Lasorne2005}.
Indeed, this rare topological phenomenon has not been addressed at all
in recent authoritative overview articles on
reaction path bifurcations~\cite{Ess2008,Rehbein2011}, 
whereas 
a rigorous definition can be found in Ref.~\citenum{Quapp2004}.

Mechanochemical ring-opening of cyclopropane derivatives 
(see Fig.~\ref{fig:reaction}) 
might be a good candidate reaction to search for
force-induced bifurcation changes in view of various surprising 
findings in the recent literature.\cite{Lenhardt2009,Lenhardt2010,%
Dopieralski2011,Akbulatov2012,Klukovich2013,Wang2014,Wollenhaupt2015,Wang2015,Wang2015b,Martinez-Craig-2015,boulatov-craig-2016,Wang-2016}
In particular, \textit{ab initio} molecular dynamics simulations at constant force 
revealed very puzzling behavior 
of {\em trans}-1,1-dichloro-2,3-dimethyl\-cyclo\-propane 
upon disrotatory ring-opening 
at finite temperatures,\cite{Dopieralski2011} namely the preferential generation
of two different diastereomers depending on the magnitude of the force
(see Fig.~3 in Ref.~\citenum{Dopieralski2011});
note that this reactant,  which is 
(2\textit{S},3\textit{S})-1,1-dichloro-2,3-dimethyl\-cyclo\-propane,
is abbreviated in the literature by 
\textit{trans}-\textit{gem}-dichlorocyclopropane 
or by {\em trans-g}DCC. 
Spurred by the findings, we explored the 
force-transformed potential energy surfaces (FT-PES)
of this 
cyclopropane derivative in more detail based on static isotensional calculations
and discovered a qualitative change in the
energetically preferred reaction channel.\cite{Wollenhaupt2015} 
Analyzing the
Intrinsic Reaction Coordinates (IRC),
we
could show that disrotatory ring-opening,
which is symmetry-allowed in the thermal limit at zero force,
switches to the conrotatory process at forces exceeding about 1.6~nN 
(being symmetry-forbidden in the thermal activation limit).\cite{Wollenhaupt2015} 
Moreover, we demonstrated that the conrotatory mechanism
does even exist down to zero force where it is symmetry-forbidden,
albeit as a high-energy reaction channel.\cite{Wollenhaupt2015}
These key computational findings~\cite{Wollenhaupt2015} 
are in line with 
a subsequent publication on 
single-molecule force spectroscopy measurements.\cite{Wang2015}  

In this work, we significantly transcend the current knowledge
by demonstrating that a system class as simple as
cyclopropane derivatives indeed features
not only downhill but also uphill bifurcations which can
be manipulated systematically by applying mechanical forces. 
The \textit{trans}-\textit{gem}-dihalocyclopropanes
being in the focus of the present study 
are able to undergo disrotatory and conrotatory ring-opening reactions
together with subsequent halogen migration to 
one or the other neighboring carbon site within the three-membered ring
as shown in Fig.~\ref{fig:reaction}.
%
%
%
%
%
%
In case of the fluorine disubstituted system it is important
to note that halogen migration corresponds to a higher-lying
reaction channel than the experimentally observed
ring-opening/closing process,\cite{Borden1994,Borden1998_exp} 
which we discuss in detail in 
Sec.~1.2 of the SI.
Ring-opening with subsequent halogen migration
leads to four product species in total
which can be classified in terms of
Woodward-Hoffmann (WH) 
allowed and forbidden electrocyclic reactions
as a result of thermal activation
together with the migration direction of the halogen. 
Upon mechanochemical activation,
it will be unveiled that this system features fundamental
topological phenomena since both contributions to the full reaction mechanism are
connected to 
uphill/downhill
bifurcations that can be tuned by external force. 
In order to set the stage, we initially investigate
the disrotatory reactions of differently substituted 
dihalocyclopropanes which are the WH allowed processes 
upon thermal activation. 
We discover that
the bifurcations deciding on the migration direction of the moving atom 
depend on the particular halogen,
and thus that these bifurcations
can indeed be manipulated by means of classic chemical substitution. 
For the total disrotatory and conrotatory reaction profile of \textit{trans}-\textit{gem}-dichlorocyclopropane
we find both types of bifurcation to be present at zero force. 
The subsequent core part of our investigation is focused on
how both the downhill and uphill bifuractions in the dichloro derivative 
can be manipulated 
by applying tensile forces to the two methyl substituents
and, in addition, reveals catastrophes at specific critical force magnitudes. 

\section{Results and Discussion}
\label{sec:Results and discussion}
\subsection{Chemical Substitution of Cyclopropanes: From Fluorine to Iodine}
\label{subsec: Chemical Substitution: From Fluorine to Iodine}
\begin{figure}
	\centering
	    \includegraphics[width=0.4\textwidth]{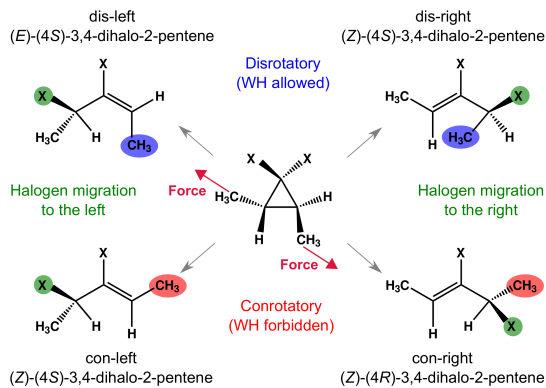}
    	\caption{%
The reactant species 
(i.e. (2\textit{S},3\textit{S})-1,1-dihalo-2,3-dimethylcyclopropanes 
a.k.a. \textit{trans}-\textit{gem}-dihalocyclopropanes
or {\em trans-g}DHC)
at the center and the corresponding four products 
in the corners are schematically depicted 
including their IUPAC nomenclature as well as
our short hand notation that is used in the present work;
see Sec.~1.1 in the SI for details regarding our nomenclature.
}
	\label{fig:reaction}
\end{figure}
\begin{figure}
	\begin{center}
	\centering
	\begin{subfigure}{0.23\textwidth}
	\centering
		\includegraphics[width=\textwidth]{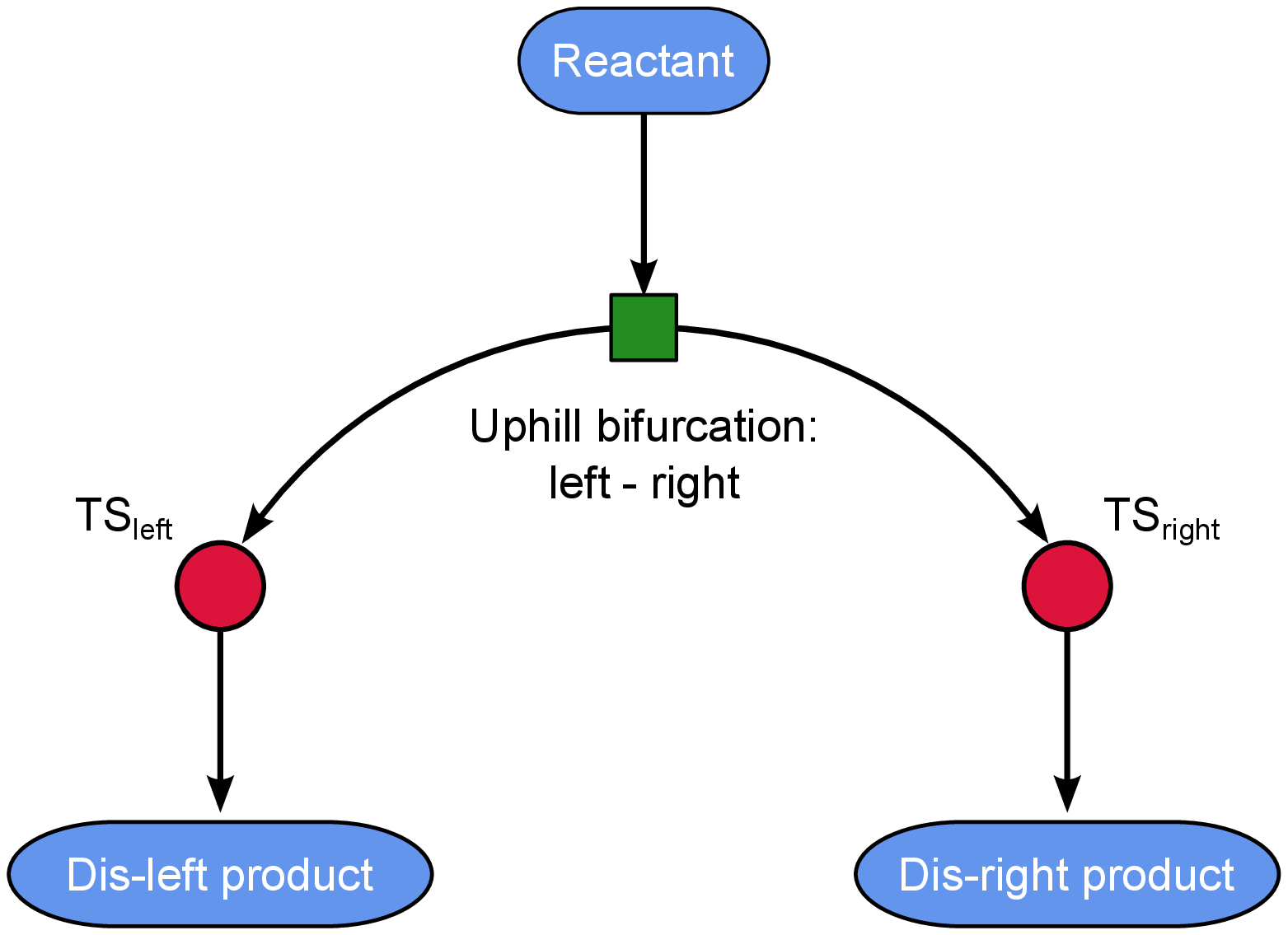}
		\subcaption{Substitution with fluorine:\\
		Uphill bifurcation.}
		\label{fig:Bifurcations_Halogens_Fluor_Uphill}
	\end{subfigure}
	\begin{subfigure}{0.23\textwidth}
	\centering
		\includegraphics[width=\textwidth]{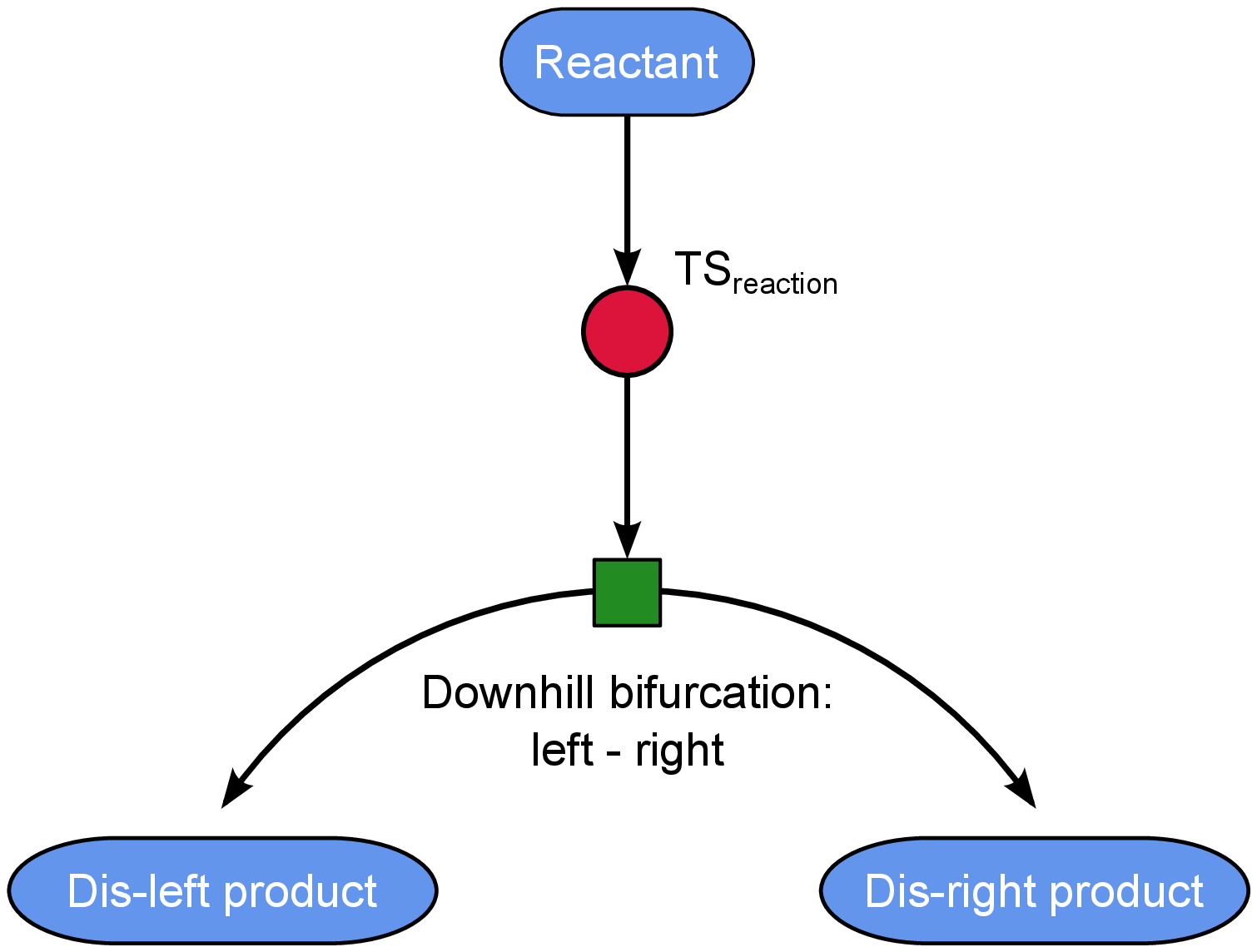}
		\subcaption{Substitution with chlorine, bromine, 
and iodine:
		Downhill bifurcation.}
		\label{fig:Bifurcations_Halogens_Cl-Br-I_Downhill}
	\end{subfigure}
    	\caption{%
Disrotatory reaction schemes depending on the dihalogen substitution 
of (2\textit{S},3\textit{S})-1,1-dihalo-2,3-dimethylcyclopropane
in the thermal reference case (corresponding to the zero-force limit).
Reactant and product minima, transition states, and bifurcations
are depicted in blue, red and green,
respectively.
		}
	\label{fig:Bifurcations_Halogens}
	\end{center}
\end{figure}

As a first step toward investigating putative force-induced topology changes
of the PES of cyclopropane derivatives, cf. Fig.~\ref{fig:reaction}, 
the bifurcation structures of the 
(2\textit{S},3\textit{S})-1,1-dihalo-2,3-dimethylcyclopropanes
are analyzed in the usual thermal limit for the fluorine 
up to iodine disubstituted species. 
It is found that the difluorinated cyclopropane derivative exhibits two 
separate disrotatory transition states, one leading to what we call the fluorine-left-product 
whereas the other one connects to the fluorine-right-product (left and right being defined 
for the migration of the halogen within
our reference frame as specified in Fig.~S1 of the SI).
%
%
%
%
%
%
These two transition states correspond to energetically 
higher-lying reaction pathways compared to the theoretically~\cite{Borden1994}
predicted 
and experimentally~\cite{Borden1998_exp} established ring-opening/closing
reaction channels of this system as detailed in Sec.~1.2 of the SI.
%
However, in the present case exactly these higher-lying reaction
pathways of the ring-opening and subsequent fluorine migration processes
are of fundamental interest in order to coherently compare 
%
%
the changes of the PES upon chemical substitution.
%
The paths to these two first-order saddle points on the PES start at the same 
stationary point representing the reactant species and initially share the identical pathway,
both structurally and energetically, when moving uphill 
(the energies along the corresponding IRCs are shown in 
Fig.~S3 of the SI).
At some specific point, however, an uphill bifurcation occurs where two distinct
pathways emerge which evolve separately {\em via} their own transition states,
TS$_{\rm left}$ and TS$_{\rm right}$,
to two different stationary states 
being the respective disrotatory products, dis-left and dis-right, 
as illustrated in Fig.~\ref{fig:Bifurcations_Halogens}(a).
This is exactly the scenario that has been discussed in the introduction
with the help of Fig.~\ref{fig:Bifurcations_Generell}(b)
and therefore adds (2\textit{S},3\textit{S})-1,1-difluoro-2,3-dimethylcyclopropane 
to the few known chemical reactions that feature an uphill bifurcation.%

In stark contrast to the difluoro species, substitution with chlorine, bromine and iodine leads 
to one common thermal disrotatory transition state, TS$_{\rm reaction}$, and a 
subsequent downhill bifurcation 
such as the one sketched in Fig.~\ref{fig:Bifurcations_Generell}(a)
that decides about the direction of the migration 
of the moving halogen atom to the dis-left and dis-right product states.

\begin{figure}
	\begin{center}
	\centering
	\begin{subfigure}{0.23\textwidth}
	\centering
		\includegraphics[width=\textwidth]{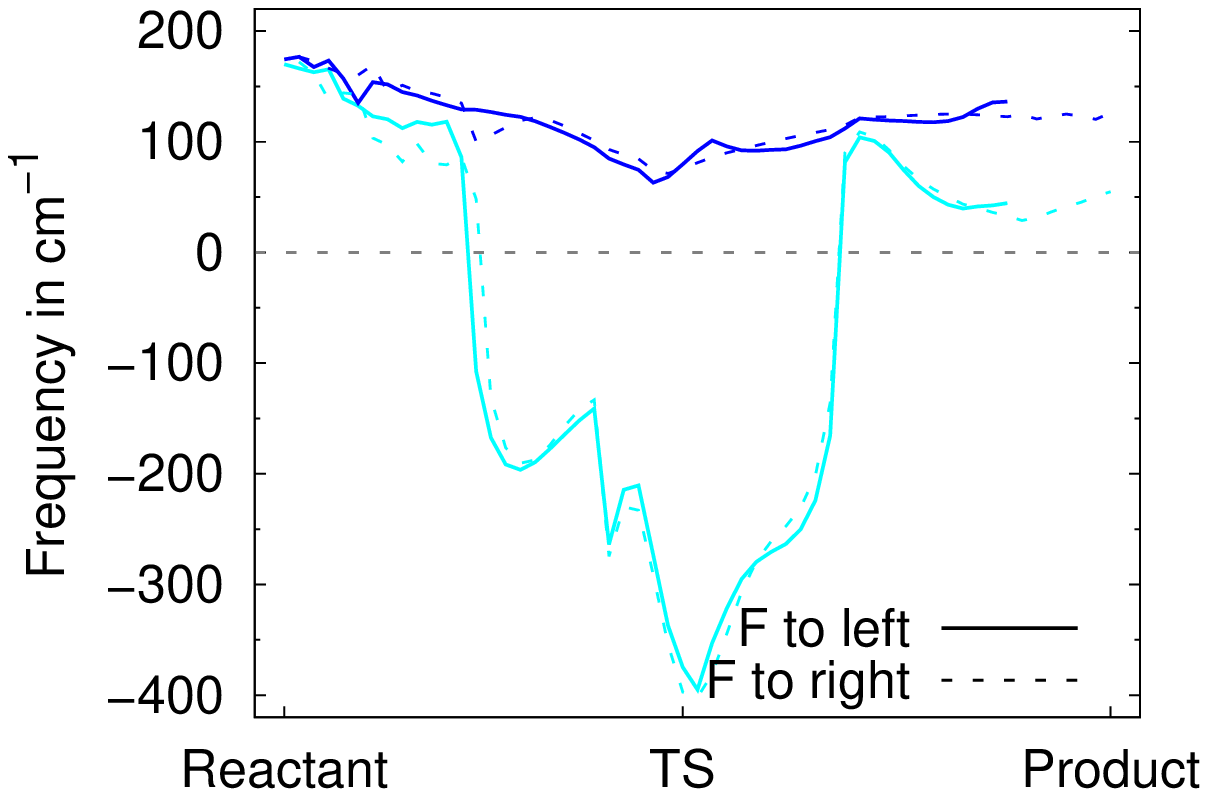}
		\subcaption{Fluorine}
	\end{subfigure}
	\begin{subfigure}{0.23\textwidth}
	\centering
		\includegraphics[width=\textwidth]{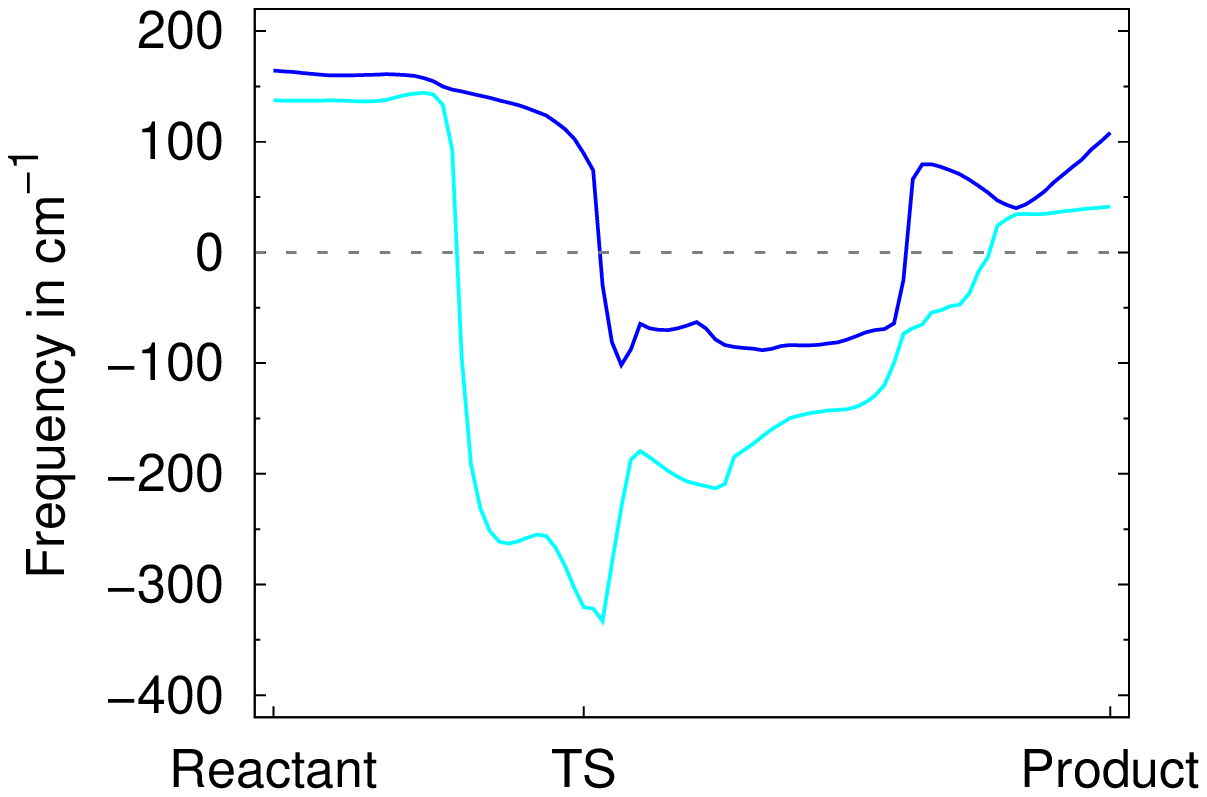}
		\subcaption{Chlorine}
	\end{subfigure}
	\begin{subfigure}{0.23\textwidth}
	\centering
		\includegraphics[width=\textwidth]{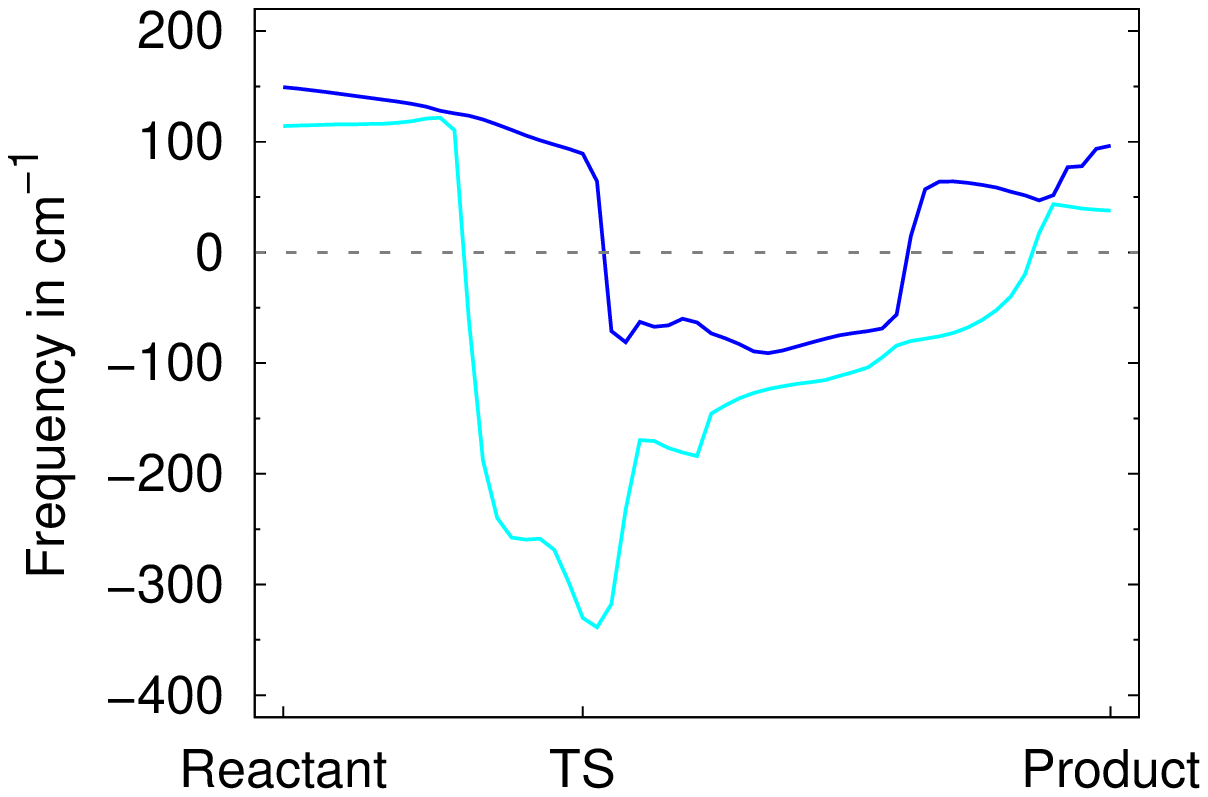}
		\subcaption{Bromine}
	\end{subfigure}
	\begin{subfigure}{0.23\textwidth}
	\centering
		\includegraphics[width=\textwidth]{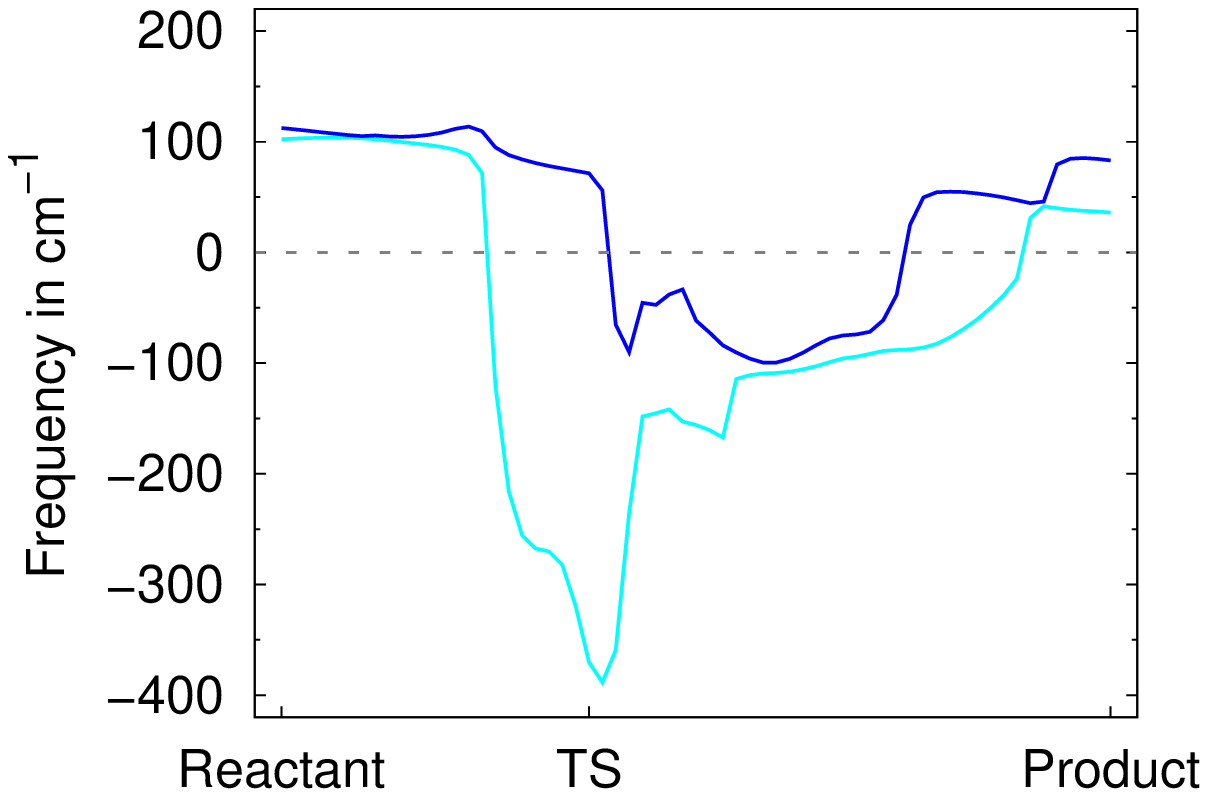}
		\subcaption{Iodine}
	\end{subfigure}
    	\caption{Frequency analysis along the IRCs for disrotatory ring-opening of 
(2\textit{S},3\textit{S})-1,1-dihalo-2,3-dimethylcyclopropane
derivatives subject to F~(in panel~a), Cl~(b), Br~(c), and I~(d) disubstitution 
in the thermal reference case, $F_0=0$~nN.
The evolution of the magnitude of the two lowest frequencies 
(negative values correspond to imaginary frequencies) 
is shown from the reactant state {\em via} the transition state to the product state. 
		}
	\label{fig:Freqs-along-IRCs_Halogens}
	\end{center}
\end{figure}

Investigating the 
evolution of particular vibrational modes along IRCs
has recently been successfully employed in order to analyze 
bifurcations~\cite{Maeda2015},
although frequency analyses have to be interpreted carefully if the
analyzed structure is not a stationary point on the PES.
(see also Sec.~1.4 in the SI).
In the present case, the evolution of the lowest 
two frequencies of the fluorine species along
the IRC corresponding to the thermal disrotatory ring-opening reaction 
depicted in Fig.~\ref{fig:Freqs-along-IRCs_Halogens}(a) 
is distinctly different from that of the chlorine~(b), bromine~(c) and iodine~(d) 
disubstituted cyclopropane derivatives (which all look quite similar). 
Only one imaginary frequency is observed for disrotatory ring-opening of the difluoro species
which is related to the fluorine-carbon stretch that
belongs to the transition state of the overall reaction
either along TS$_{\rm left}$ or TS$_{\rm right}$ toward the
dis-left and dis-right product, respectively. 
The appearance of a single imaginary frequency along a reaction path that bifurcates, 
which is left/right fluorine migration in the present case
as indicated using solid/dashed lines in Fig.~\ref{fig:Freqs-along-IRCs_Halogens}(a), 
is a characteristic feature of the above-mentioned uphill bifurcation
being in accord with the topology of Fig.~\ref{fig:Bifurcations_Generell}(b). 

In comparison to the fluorine case, the IRCs of the three other halogen derivatives 
feature more than one imaginary frequency. 
The same halogen-carbon stretch as in the fluorine case 
becomes most prominently imaginary when reaching the transition state, followed by
one additional imaginary frequency that is directly related to the
bifurcation that sets in after having passed TS$_{\rm reaction}$
when moving toward the products.
It is essentially a bending vibration that is mainly related to
the halogen left/right migration leading directly to the
dis-left/right products
and also describes the
pathway that directly interconverts the dis-left and dis-right products
via TS$_{\rm 1-2}$;
a detailed decomposition of this frequency analysis along the reaction paths 
for the chlorine case even for the force transformed PES can be found in 
Sec.~1.4.2 of the SI.
This mode softening scenario agrees nicely with what is expected for
a downhill bifurcation according to the generic PES in 
Fig.~\ref{fig:Bifurcations_Generell}(a). 

It is concluded that the difluorinated derivative of 
(2\textit{S},3\textit{S})-1,1-dihalo-2,3-dimethyl\-cyclo\-propanes
features one of the rarely observed uphill bifurcations
upon thermally activated ring-opening
followed by fluorine migration,
while the dichlorine
species is found to be the first one in this halogen homologous series 
that is characterized by a downhill bifurcation instead.

\subsection{Mechanochemistry of 
Dichlorocyclopropane:
Force-induced Topology Changes}
\label{subsec: Mechanical forces on / Mechanochemistry of Chlorine-Derivative: Reactant regime / part}

What has been unveiled in the previous section 
provokes the question if it might be possible to manipulate the
bifurcation scenario of the dichlorocyclopropane species
upon applying tensile stress with the aim to systematically
influence the bifurcations.
The remainder of this investigation will focus on that question
by discussing the behavior of
(2\textit{S},3\textit{S})-1,1-dichloro-2,3-dimethylcyclopropane
as a function of constant force.
Within this section, the global changes of the Born-Oppenheimer
PES upon its force-transformation will be discussed, culminating
in a topological catastrophe 
when reaching a specific critical force $F_0^{\rm crit}$.
Afterwards, the topological details in terms of how the uphill and downhill
bifurcations and transition states shift as a function of mechanical force
will be addressed separately for the respective parts of the FT-PES
in the subsequent sections. 

A broad understanding of the reaction behavior of this cylcopropane derivative under
the influence of finite tensile forces has been obtained earlier
based on dynamical explorations of wider regions of its energy landscape.\cite{Dopieralski2011}
This included extensive mappings of multi-dimensional 
free energy surfaces for disrotatory 
ring-opening based on isotensional {\em ab initio} 
metadynamics as well as corresponding 
trajectory shooting simulations at a set of finite forces.
Our subsequent static analysis of not only the disrotatory 
pathway as before,\cite{Dopieralski2011} but also of the 
conrotatory ring-opening along the corresponding IRCs
as a function of the external force has brought deeper insights into the
evolution of the electronic structure along these distinct reaction
channels.\cite{Wollenhaupt2015}

\begin{figure*}[htb]
	\centering
\includegraphics[width=0.65\textwidth]{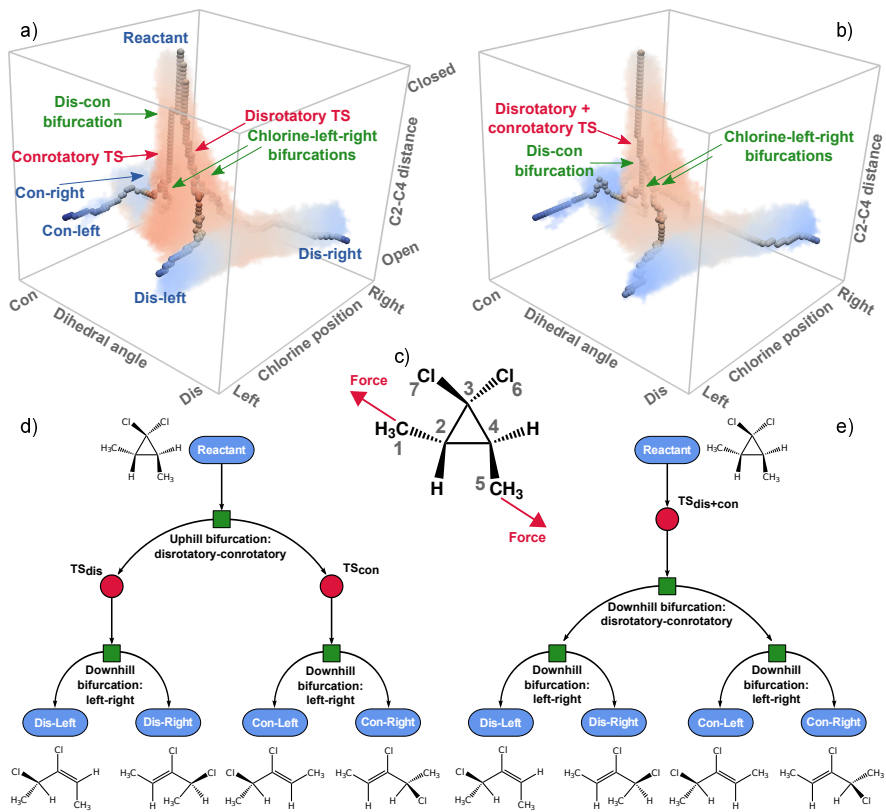}
    	\caption{%
Force-transformed effective potential energy surfaces (FT-ePES)
at two different constant 
forces of $F_{0}=1.0$~nN and $F_{0}=2.0$~nN in panel~(a) and~(b), respectively,
depicted in the reaction subspace spanned by the three selected collective variables (see text)
as indicated.
In addition, high/low energy parts of the FT-ePES are encoded using red/blue colors. 
The corresponding Dijkstra paths (see Sec.~1.6 in the SI for details)
are included using a necklace representation.
Panel~(c) introduces the stereochemical formula of the
(2\textit{S},3\textit{S})-1,1-dichloro-2,3-dimethylcyclopropane 
reactant, its atom numbering, and also indicates how the constant colinear force $F_0$ 
acts on the two methyl groups. 
Panels~(d) and~(e) represent the two topologically distinct 
reaction schemes valid for zero (and moderate) forces in~(d) and large
forces in~(e) corresponding to the FT-ePES of panel~(a) and~(b), respectively,
and also display the stereochemical formulae of all products.
}
	\label{fig:ReactionSchemes}
\end{figure*}
In what follows, we combine these approaches in the sense that we now use
isotensional {\em ab initio} molecular dynamics at several constant forces 
(within the EFEI approach as explained in the SI)
in order to sample 
a vast ensemble of configurations with corresponding electronic energies.
This allows us to generate the FT-PES as a function of force in 
a low-dimensional subspace which, in a second
step, will be analyzed in terms of bifurcations and a 
catastrophe after adopting a static topology viewpoint;
the generation of the FT-PES based on isotensional trajectories at a set
of constant forces is explained in Sec.~1.5 of the SI.
Three key structural parameters or collective variables (CVs) 
have been identified of being able to describe the important aspects of the processes 
in a reduced reaction subspace, see  Fig.~\ref{fig:ReactionSchemes}, 
rather than considering all internal degrees of freedom in full dimensionality. 
As the ring-opening and thus the bond breaking of the carbon-carbon bond 
within the three-membered cyclopropane ring constitutes a major part 
of the reaction, the corresponding distance \mbox{(C2--C4)} 
has been chosen as the first variable, CV1. 
Next, the dihedral angle of what we call the right methyl group is used as CV2 
to discriminate between disrotatory and conrotatory trajectories.
Last but not least, in order to differentiate between the left and right products, 
the relative position of the migrating chlorine atom, which is Cl7, is employed as 
our third dimension, CV3. 
Therefore, the resulting FT-PES is not represented in the full-dimensional
space defined in terms of all internal degrees of freedom, but in the 
three-dimensional (3D) reaction subspace spanned by the chosen 
collective variables, i.e. PES(CV1,CV2,CV3), which is why we call it the
(force-transformed) ``effective potential energy surface'' \mbox{(FT-) ePES}. 
This dimensionality reduction from 
39 down to only~3 variables turns out to be crucial not only to analyze, but also to
visualize the force-induced changes of the bifurcations.  
On the FT-ePES, all four reaction paths from the reactant
to the four possible products have been located by applying
the Dijkstra algorithm,\cite{Dijkstra1959}
thus providing what we call Dijkstra paths.
Note that standard approaches to determine reaction pathways
fail to map bifurcations which are usually not stationary points.
The Dijkstra algorithm, in detail described in
Sec.~1.6.1 of the SI, is therefore crucial to fully understand the
changes of the ePES with respect to force transformation.

Depending on the magnitude of the applied external force, we have found two 
topologically distinct reaction schemes that can connect the same reactant state 
with the four products.
These scenarios correspond to uphill and downhill bifurcations 
according to panels~(a,c) and (b,e) in Fig.~\ref{fig:ReactionSchemes}
which are separated by a catastrophe that is triggered when using the
magnitude of the external force as the control parameter in the spirit of
catastrophe theory.\cite{Gilmore1981} 
For clarity, we will first describe and explain the reaction course that 
is observed at zero force, i.e. in limit of a thermally activated process,
which is a scenario that is still topologically valid at moderate finite forces.
As depicted in the left part of Fig.~\ref{fig:ReactionSchemes}, the processes
leading to the four products 
share initially the same pathway on the FT-ePES. 
At some point, however, an uphill bifurcation deciding between disrotatory and 
conrotatory ring-opening occurs 
(denoted by ``dis-con bifurcation'' in panel~(a) of Fig.~\ref{fig:ReactionSchemes}) 
where the common path splits into two distinct ascending reaction channels 
before any transition state is reached.
After having passed this bifurcation that is still on the reactant side of the energy landscape, 
each of the paths climbs further up in energy and reaches its own transition state,
denoted by con- and disrotatory TS in panel~(a);
note that irrespective of the occurrence of (uphill/downhill) bifurcations
along a specific path the transition state remains always the
(stationary first-order saddle) point of highest energy along that pathway
whereas bifurcations do usually not correspond to stationary points 
and thus cannot be detected using the standard geometry optimizers in quantum chemistry packages. 
Only from the transition state
on do the pathways descend in energy toward the products.
One of the transition states corresponds to the disrotatory ring-opening reaction, 
whereas at the other the methyl groups rotate in the same fashion thus 
following a conrotatory mechanism.

Overall, we observe the following scenario at zero up to moderate forces,
see Fig.~\ref{fig:ReactionSchemes}(d). 
All four paths are degenerate before reaching the uphill bifurcation in the
ascending part of the FT-ePES where they split into two pairs which remain
degenerate all the way up to their transition states.
After having passed their own transition state, thus being now on the descending
side of the FT-ePES, 
they remain degenerate for a while before reaching the
downhill bifurcation, where they finally split into left and right products.
Hence, the 
four respective product channels represent a combination of
con/disrotatory ring-opening and left/right chlorine migration
that involve 
uphill and downhill bifurcations, respectively.

At forces higher than a specific critical value 
(denoted as $F_{0}^{\rm{crit}}$,  see next section for its quantification), 
the reaction scenario changes significantly as supported by the
topology changes of the FT-ePES, see right panels of Fig.~\ref{fig:ReactionSchemes}.
In this high-force regime,  again all four paths start in the reactant minimum
but remain degenerate and thus do not bifurcate before they reach 
their common transition state.
In contrast to the reaction route at sub-critical forces including the thermal activation limit, 
the FT-ePES in Fig.~\ref{fig:ReactionSchemes}(b) 
does not contain any uphill bifurcation since a
common transition state is reached directly 
after the ascending side of the energy landscape. 
After this joint transition state, however, a first downhill bifurcation 
can be found which leads to the splitting of the common pathway
into two reaction channels 
(each being two-fold degenerate) 
corresponding to disrotatory and conrotatory ring-opening.
Subsequently, yet another pair of downhill bifurcations occurs which decide on the left/right 
movement of the migrating chlorine atom (Cl7) after
either con- or disrotatory ring-opening, and thus lift the remaining
degeneracy by splitting the two paths for a second time. 

\subsection{Mechanochemistry of 
Dichlorocyclopropane: 
Reactant Regime} 
\begin{figure}
    \centering
	\begin{subfigure}{0.23\textwidth}
	\centering
	    \includegraphics[width=\textwidth]{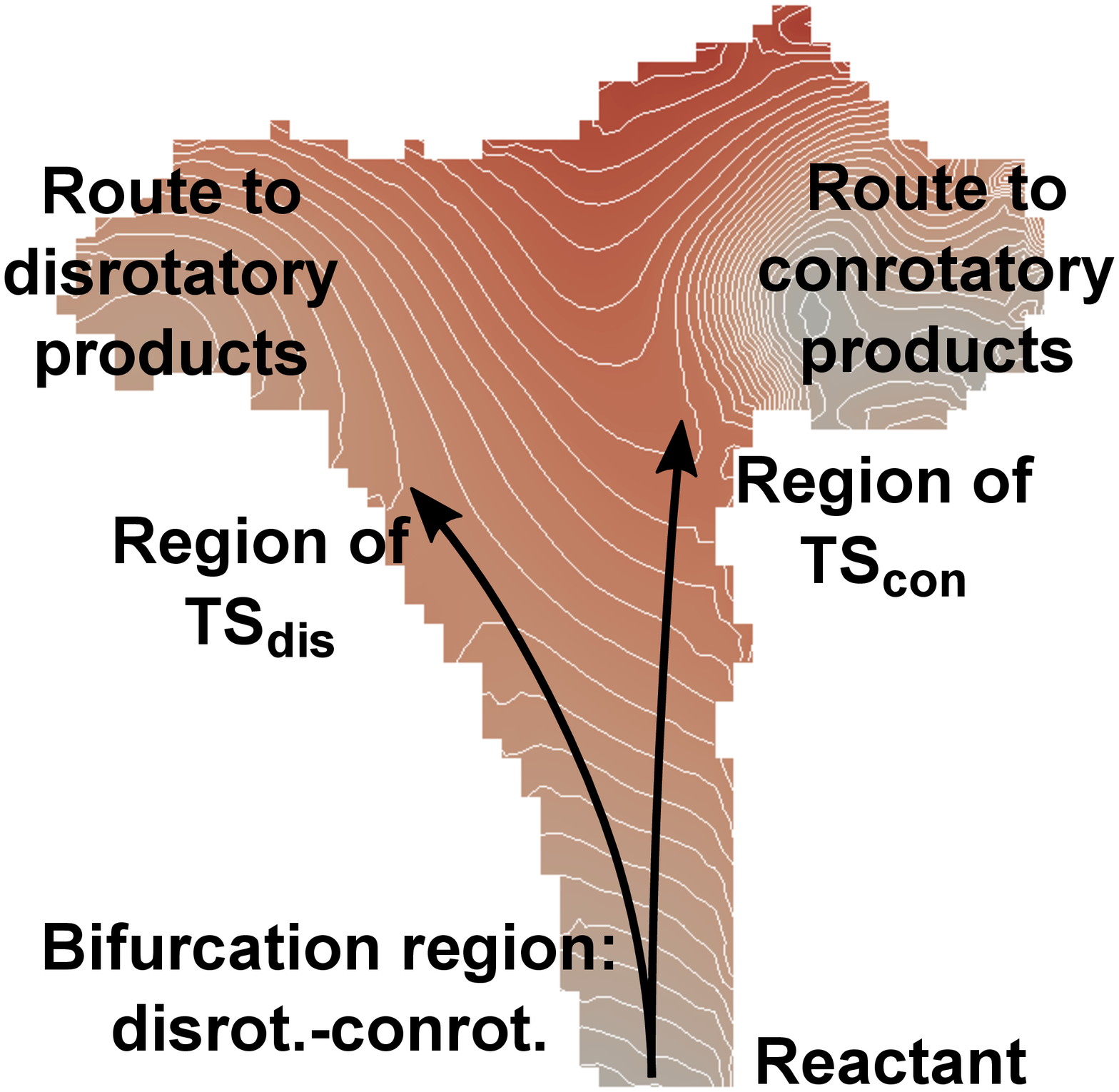}
		\subcaption{$F_{0}$ = 0.0 nN}
	\end{subfigure}
	\begin{subfigure}{0.23\textwidth}
	\centering
	    \includegraphics[width=\textwidth]{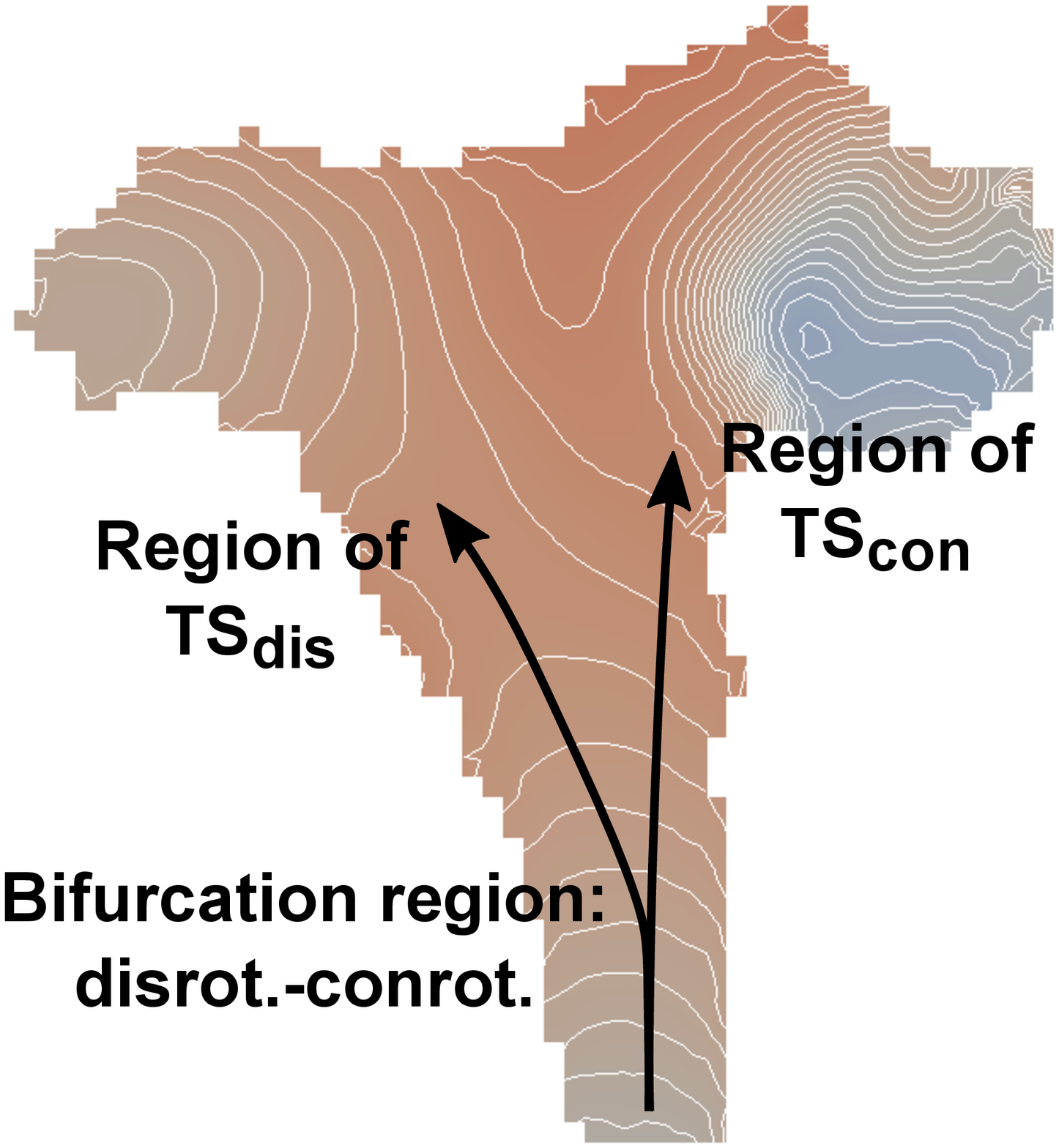}
		\subcaption{$F_{0}$ = 1.0 nN}
	\end{subfigure}
	\begin{subfigure}{0.23\textwidth}
	\centering
	    \includegraphics[width=\textwidth]{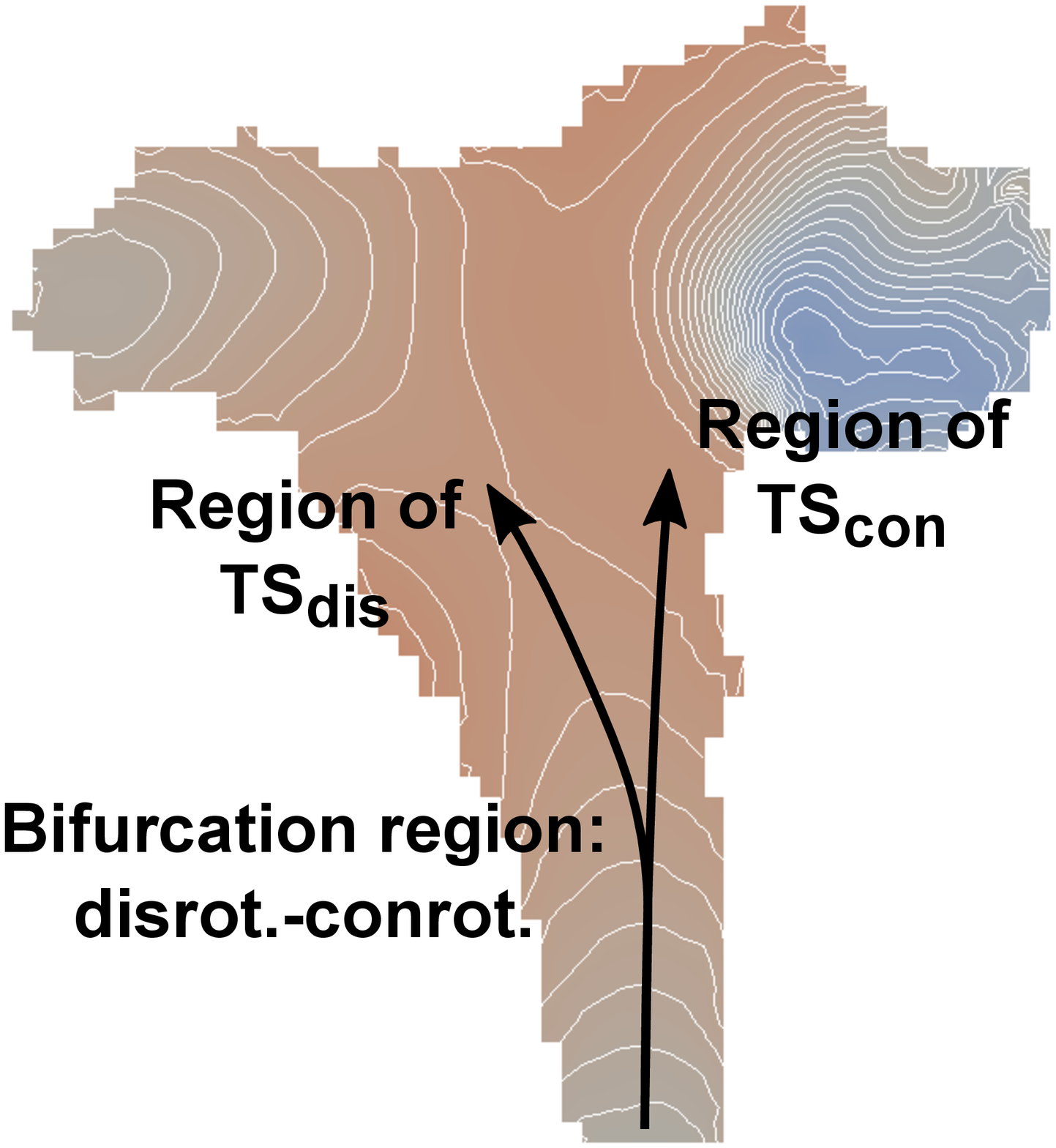}
		\subcaption{$F_{0}$ = 1.5 nN}
	\end{subfigure}
	\begin{subfigure}{0.23\textwidth}
	\centering
	    \includegraphics[width=\textwidth]{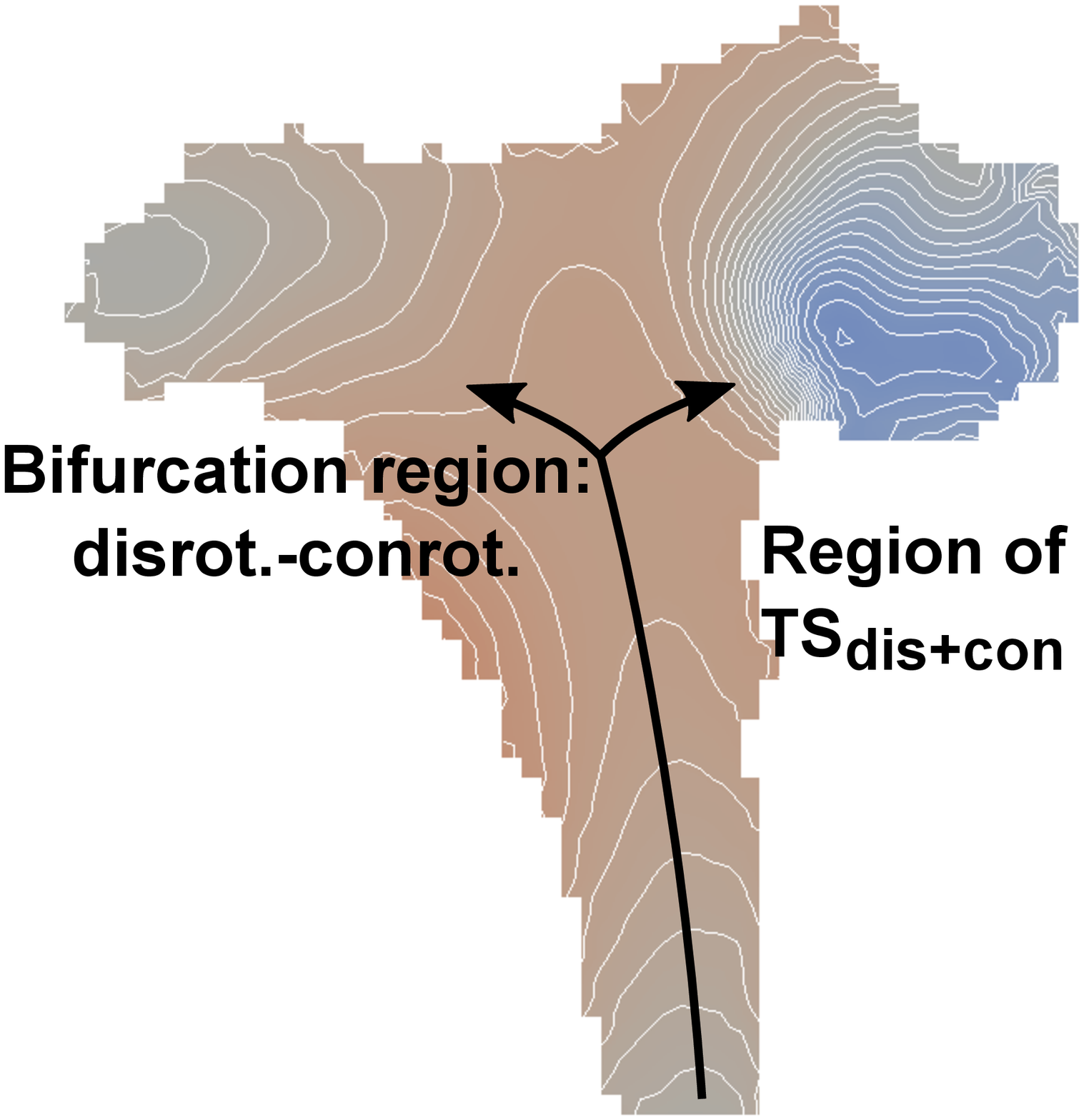}
		\subcaption{$F_{0}$ = 2.0 nN}
	\end{subfigure}
	\begin{subfigure}{0.23\textwidth}
	\centering
	    \includegraphics[width=\textwidth]{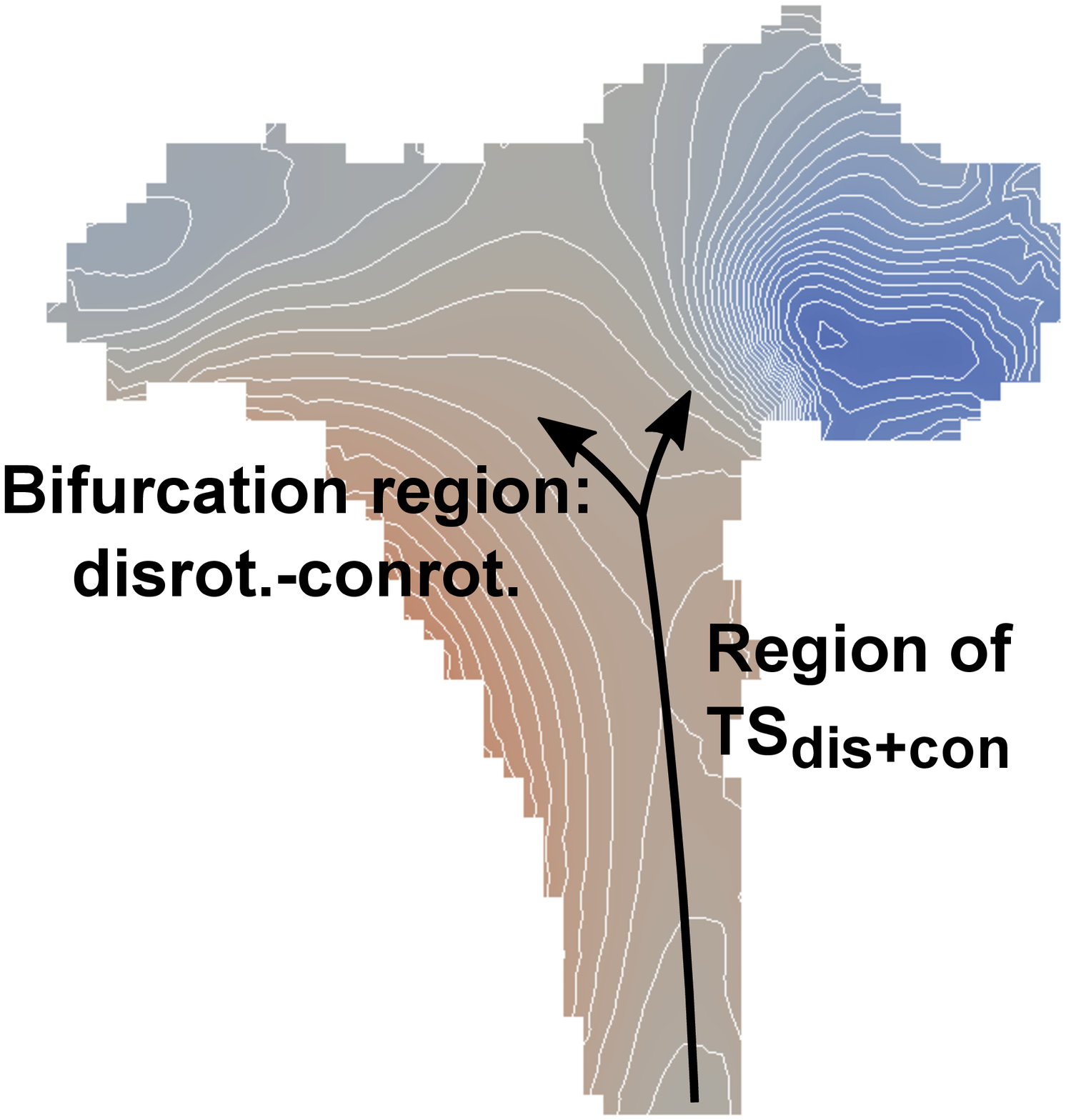}
		\subcaption{$F_{0}$ = 3.0 nN}
	\end{subfigure}
    \caption{Schematic illustration of the disrotatory and conrotatory 
Dijkstra paths on a set of 2D-slices of the 3D FT-ePES obtained at 
different constant forces from the thermally activated limit at
0~nN in panel~(a) up to 3~nN in (e).
The $x$-axis represents the dihedral angle of the methyl group (CV2) whereas the
$y$-axis corresponds to the length of the breaking carbon-carbon bond (CV1). 
The reaction paths 
lead from the reactant region (bottom part of each panel) to the region
of transition state/s  (central part) and finally toward the 
dis- and conrotatory products (left and right upper parts, respectively);
the arrow heads indicate the transition states whereas the branching
points correspond to the bifurcations.  
The reactant energy is set to \SI{0}{\kilo\calorie\per\mole} for each FT-ePES 
and the depicted energy contour lines are separated by an equidistant 
spacing of \SI{2}{\kilo\calorie\per\mol}
where red encodes increasing and blue decreasing energies compared to the reactant energy.
	}
	\label{fig:2D-Slices-Reactant_Energies_+_marked_Paths}
\end{figure}
The quasi-3D representations of Fig.~\ref{fig:ReactionSchemes}(a) and (b)
are useful to visualize and discuss the global topology corresponding 
to the sub- and super-critical
reaction scenarios, respectively, as they take place on the FT-ePES. 
Yet, only 
their reduction to specifically selected two-dimensional slices 
allows one to gain deeper insights as depicted in
Fig.~\ref{fig:2D-Slices-Reactant_Energies_+_marked_Paths}
for several constant forces from the thermal activation limit in panel~(a)
up to 3~nN in (e). 
These 2D-slices have been optimized in such a way that they cover the relevant parts 
of the 3D-PES with a focus on those regions where the bifurcations take place;
one coordinate is essentially the breaking carbon-carbon bond and
thus describes the first part of the reaction progress whereas the
second coordinate is mostly the methyl dihedral and thus discriminates
between the con- and disrotatory parts of the FT-ePES. 
The energy barriers for the reaction, being the difference between the reactant and the transition
state regions, decrease with increasing mechanical forces 
(see Sec.~1.6.2 and~1.6.3 in the SI for reaction paths on the FT-ePES).

The force-dependent Dijkstra paths corresponding to the disrotatory and conrotatory ring-opening 
are schematically sketched on these 2D-slices using lines and arrows as 
collected in Fig.~\ref{fig:2D-Slices-Reactant_Energies_+_marked_Paths};
see Sec.~1.6.1 in the SI for details on computing the Dijkstra paths numerically. 
At zero force, i.e. in the thermal reference case~(a), all paths start at the 
same reactant but diverge readily afterwards. 
The left paths lead to the disrotatory side of the FT-ePES while the right paths 
connect the reactant to the conrotatory products;
note that the left/right migration that decides about the final product
after the ring-opening cannot be seen in this subspace of coordinates.
As seen from the pattern of the contour lines, the initial part of the reaction channel 
is quite wide, which provides the possibility for an early divergence of the 
two paths that eventually lead to the dis- and conrotatory products. 
Clearly, the bifurcation between disrotatory and conrotatory ring-opening is
positioned on the part of the FT-ePES with ascending energy, i.e. 
long before the two corresponding transition states are 
reached and, therefore, represents an uphill bifurcation.
The two distinct transition states for the dis- and conrotatory reactions,
TS$_\text{dis}$ and TS$_\text{con}$,
are located in the central left and right region about where the
arrow heads are. 
Moreover, the two transition states are clearly separated from each
other by an energy barrier which keeps these two reaction channels apart.
This zero-force scenario is qualitatively valid up to a force of
roughly 1.5~nN as shown in panel~(c).

With increasing force, however, 
in the vicinity of the reactant, the reaction channel
gets more narrow, which can be seen from the changing shape of the 
contour lines in the lower part moving from (a) to (b) to (c). 
More importantly, the two paths leading eventually to the dis- and conrotatory processes
are found to increasingly share the same part of the FT-ePES,
which graphically shifts the bifuraction region upwards in the graphs
upon increasing the force. 
Concurrently, the disrotatory and conrotatory transition states 
approach each other both structurally and energetically, 
implying that the energy barrier in between them 
gets systematically suppressed. 
The significant qualitative change with respect to the 
zero- to moderate-force scenario can most clearly
be detected at a force of 3~nN, see 
Fig.~\ref{fig:2D-Slices-Reactant_Energies_+_marked_Paths}(e).
Now, the bifurcation into dis- and conrotatory ring-opening pathways
occurs clearly after a common transition state TS$_\text{reation}$
which is shared by the two distinct ring-opening processes.
This implies that the bifurcation into dis- and conrotatory ring-openings
occurs thereafter and, thus, that it
is a downhill bifurcation at such sufficiently large forces. 
In effect, the uphill bifurcation in the thermally activated reference
case has been turned into a downhill bifurcation 
and thus has been shifted from the reactant (energy ascending) side 
of the energy landscape to the product (descending) side
upon mechanochemical activation. 
In concert with this 
force-induced
shift, the two formerly separated dis- and conrotatory transition states 
become confluent at sufficiently large forces and thus generate a single transition state
that is common to both ring-opening reactions;
note that the subsequent left/right bifurcation related to chlorine migration 
remains on the downhill (product) side 
irrespective of the force as will be discussed later. 
All these phenomena are already present at 2~nN in~(d).

\begin{figure*}
	\centering
	\includegraphics[width=0.9\textwidth]{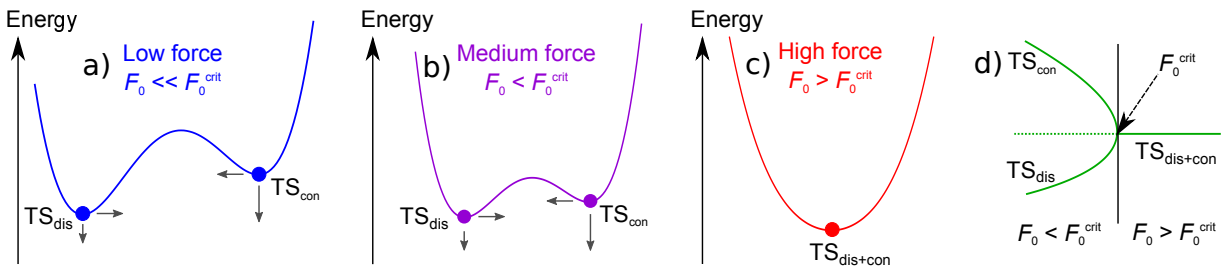}
    \caption{
Schematic representation of the force-induced distortions of the 
potential energy surface (FT-PES) in the region around the disrotatory 
($\textrm{TS}_{\textrm{dis}}$) and conrotatory ($\textrm{TS}_{\textrm{con}}$) 
transition states. 
The cut through the FT-PESs is chosen to be 
perpendicular to the pathways {\em via} both $\textrm{TS}_{\textrm{dis}}$ 
and $\textrm{TS}_{\textrm{con}}$ which implies that the transition states
appear as minima along the $x$-axis in this scheme
whereas the central barrier is the one that energetically
separates the two distinct pathways.
Increasing the magnitude of the external force $F_0$ transforms the topology of the energy landscape
from the one characterizing the sub-critical regime sketched in~(a) and also in~(b) 
to that in the super-critical limit depicted in~(c). 
The rightmost diagram~(d) visualizes as a function of the control parameter $F_0$
the change of topology upon increasing the force beyond its
critical value, $F_0 = F_{0}^{\rm{crit}}$, where the 
topological catastrophe occurs. 
}
    \label{fig:TS_Vereinigung-Schema}
\end{figure*}
Based on this topological analysis, 
we conclude from the distortion of the FT-ePES that the catastrophe, which switches 
the topology from the reaction scenario valid in the thermal limit up to moderate forces 
(see Fig.~\ref{fig:ReactionSchemes}(a) and (d))
to the large-force scenario
(see Fig.~\ref{fig:ReactionSchemes}(b) and (e)), 
occurs at a critical force $F_0^{\rm crit}$ that is bracketed by
1.5 and 2.0~nN.
However, being close to the topological catastrophe, that part of
the FT-ePES in panel~(d), which covers the confluence 
of the two distinct transition states
and the concurrent conversion of the uphill into a downhill
bifurcation which needs to shift onto the other side of the new joint transition
state, is very flat close to the critical force.
This is borne out by analyzing the force-transformed energy profiles 
on the FT-ePES along the two Dijkstra paths corresponding to the 
dis- and conrotatory ring-opening processes,
see Fig.~S8(a) of the SI.
As one can see, the conrotatory curve only slowly approaches
the lower-lying disrotatory one upon increasing $F_0$ such that the energy difference 
between the two processes eventually becomes 
insignificant before reaching $F_0^{\rm crit}$.
This energy difference is $\approx 1.2$~kcal/mol at $F_0  = 1.5$~nN
and is on the order of only 0.1~kcal/mol at 1.8 and 1.9~nN
before it vanishes at 2.0~nN, 
recall that the thermal energy, $k_{\rm B}T$, corresponds 
to $\approx 0.6$~kcal/mol at room temperature. 
Moreover, it is difficult
to fully quantitatively analyze the catastrophe given the numerical inaccuracies
due to the \textit{ab initio} molecular dynamics sampling of the 
FT-PES in all 39~internal coordinates which underlies the construction
of the coarse-grained effective energy landscape, FT-ePES, in terms of only 3~collective variables.
Last but not least, it is the full-dimensional FT-PES (and not the
reduced-dimensionality treatment in terms of the FT-ePES that we
introduced in order to access the rather complex 
bifurcation topology and its change at a function of force) which 
determines $F_{0}^{\rm{crit}}$.
The former has been analyzed previously~\cite{Wollenhaupt2015}
in terms of IRCs based on static isotensional quantum chemical calculations 
and is reproduced in Fig.~S8(b) of the SI. 
As found for the Dijkstra paths in panel~(a), the IRCs on the FT-PES
start to come close at around 1.5~nN,
where the conrotatory pathway is about 0.8~kcal/mol above the disrotatory one, 
before they essentially meet around 1.6~nN (where the corresponding energy difference 
is decreased to only $\approx 0.3$~kcal/mol whereas no
disrotatory transition state could be optimized at 1.7~nN and beyond). 
Based on this discussion, we conclude that the topological catastrophe
occurs at a critical force close to $F_{0}^{\rm{crit}}=1.6$~nN. 

\subsection{Analysis of the Catastrophe on the Force-transformed PES}
\label{subsec: D)}
With the help of catastrophe theory\cite{Gilmore1981}, the 
shift and the confluence of the disrotatory and conrotatory transition states 
of the dicholorocyclopropane species 
as a result of applying constant mechanical forces 
can be viewed as a topological problem~\cite{Ribas-Arino2009}, 
as already computationally exploited by us~\cite{Ribas-Arino2010}
quite some time back 
for another mechanochemical reaction. 
The key idea is that the systematic distortion of the usual Born-Oppenheimer PES
as a result of its force transformation that yields the FT-PES
can not only lead to quantitative 
changes for instance of activation or reaction energies,
but can also induce qualitative changes of the FT-PES compared to the 
zero force (Born-Oppenheimer) PES; 
see Sec.~4.3 in Ref.~\citenum{Ribas-Arino2012} for a review of these concepts.
These force-induced qualitative changes of the energy landscape
correspond to a switching between different topologies
of the FT-PES at some specific force value, $F_0^{\rm crit}$, thus the magnitude 
of the applied force is identified to be the ``control parameter''
in the language of catastrophe theory,\cite{Ribas-Arino2010}
whereas the catastrophe itself is the sudden change 
of topology upon reaching $F_0^{\rm crit}$.

The situation that we disclosed herein 
for dichlorocyclopropane is visualized with the help of 
Fig.~\ref{fig:TS_Vereinigung-Schema} in terms of simplified
cuts through the FT-PES. 
At forces smaller than $F_{0}^{\rm{crit}}$, an uphill bifurcation
leads to separate pathways for dis- and conrotatory ring-opening
to which the two distinct transition states that are shown
in panel~(a), i.e. TS$_{\rm dis}$ and TS$_{\rm con}$, correspond to. 
Increasing the mechanical force $F_0$ leads to systematic shifts of
TS$_{\rm dis}$ and TS$_{\rm con}$ in both energy and 
relative position within the energy landscape 
as qualitatively revealed when comparing panel~(b) to~(a). 
The activation energy associated to both pathways
is gradually lowered upon
increasing the applied external force,
which is the usual mechanochemical activation 
due to suppressing the activation barrier w.r.t. the initial state {\em at the same force}.
Thus, both TSs get lower in energy (relative to the
reactant state serving as the reference) for larger $F_0$,
but the conrotatory pathway is more susceptible to force
which implies that the energy of TS$_{\rm con}$ decreases
more quickly compared to TS$_{\rm dis}$
as symbolized by the length of the 
downward-pointing vertical arrows in panels~(a) and~(b). 
Concurrent to the lowering of the relative energy is 
also a confluence in configuration space, i.e. the two TSs move toward
each other upon increasing $F_0$
as encoded by the two horizontal arrows in these panels. 
At forces greater than $F_{0}^{\rm{crit}}$, the two individual transition states have merged 
as depicted in (c) of Fig.~\ref{fig:TS_Vereinigung-Schema}
and form one joint transition state ($\textrm{TS}_{\textrm{dis+con}}$)
that is common to both reaction pathways (which eventually lead to dis- and conrotatory
ring-opening after having passed the resulting downhill bifurcation, {\em vide infra}). 
Once TS$_{\rm con}$ and TS$_{\rm dis}$ are confluent and
$\textrm{TS}_{\textrm{dis+con}}$ has emerged, the 
two distinct uphill dis/conrotatory pathways including the 
underlying uphill bifurcation vanish as well. 
The resulting (local) cusp catastrophe occurs right at $F_0 = F_{0}^{\rm{crit}}$
as visualized in panel~(d) of Fig.~\ref{fig:TS_Vereinigung-Schema}
as a function of the continuous control parameter $F_0$.

\subsection{Mechanochemistry of
Dichlorocyclopropane:
Product Regime} 
\label{subsubsec: MC on Cl-C3: Product part}

\begin{figure}
    \centering
	\begin{subfigure}{0.23\textwidth}
	\centering
	    \includegraphics[width=\textwidth]{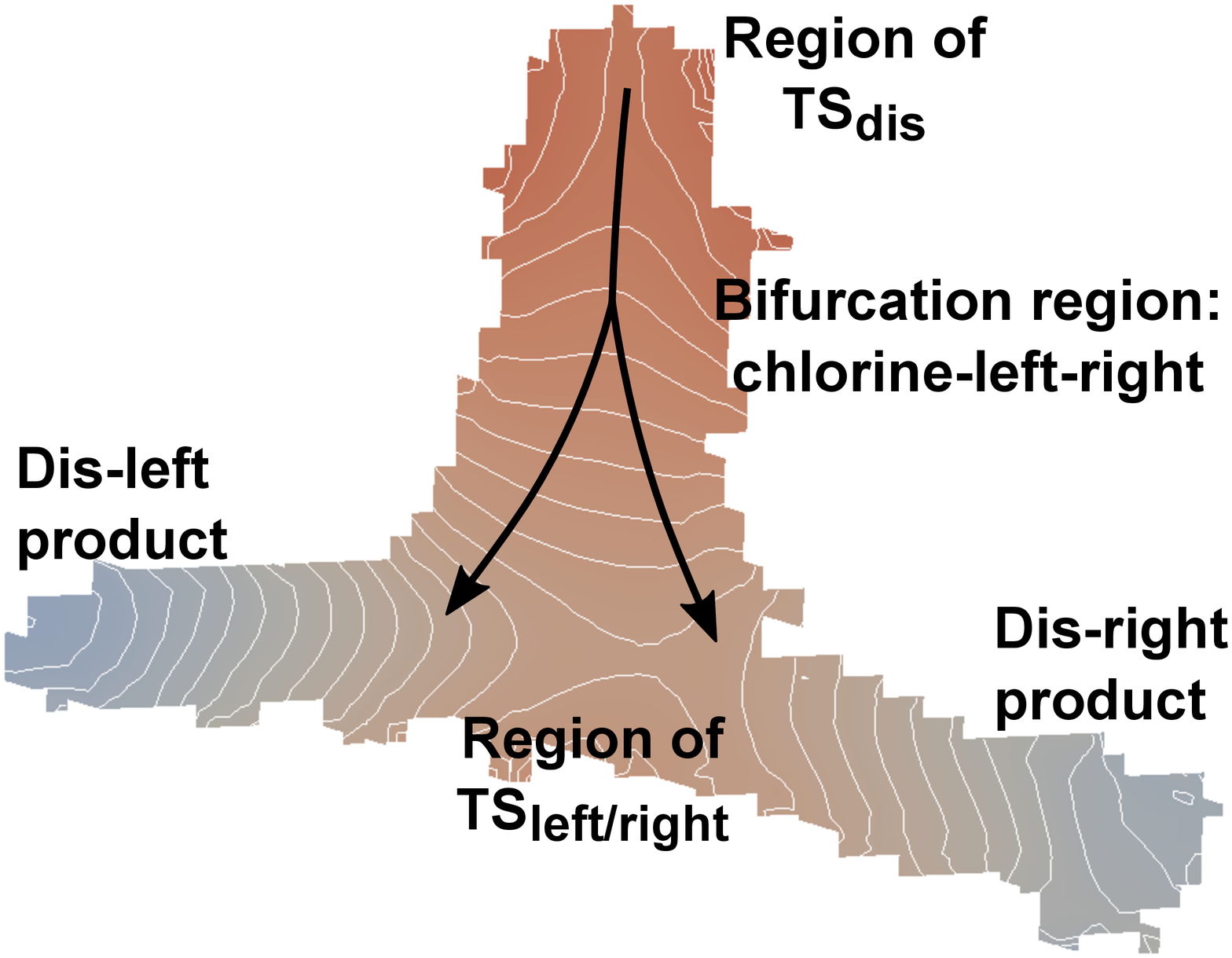}
		\caption{$F_{0} = $ 0.0~nN}
		\label{fig:PES_DisBifurcation_0.00}
	\end{subfigure}
\hfill
	\begin{subfigure}{0.23\textwidth}
	\centering
	    \includegraphics[width=\textwidth]{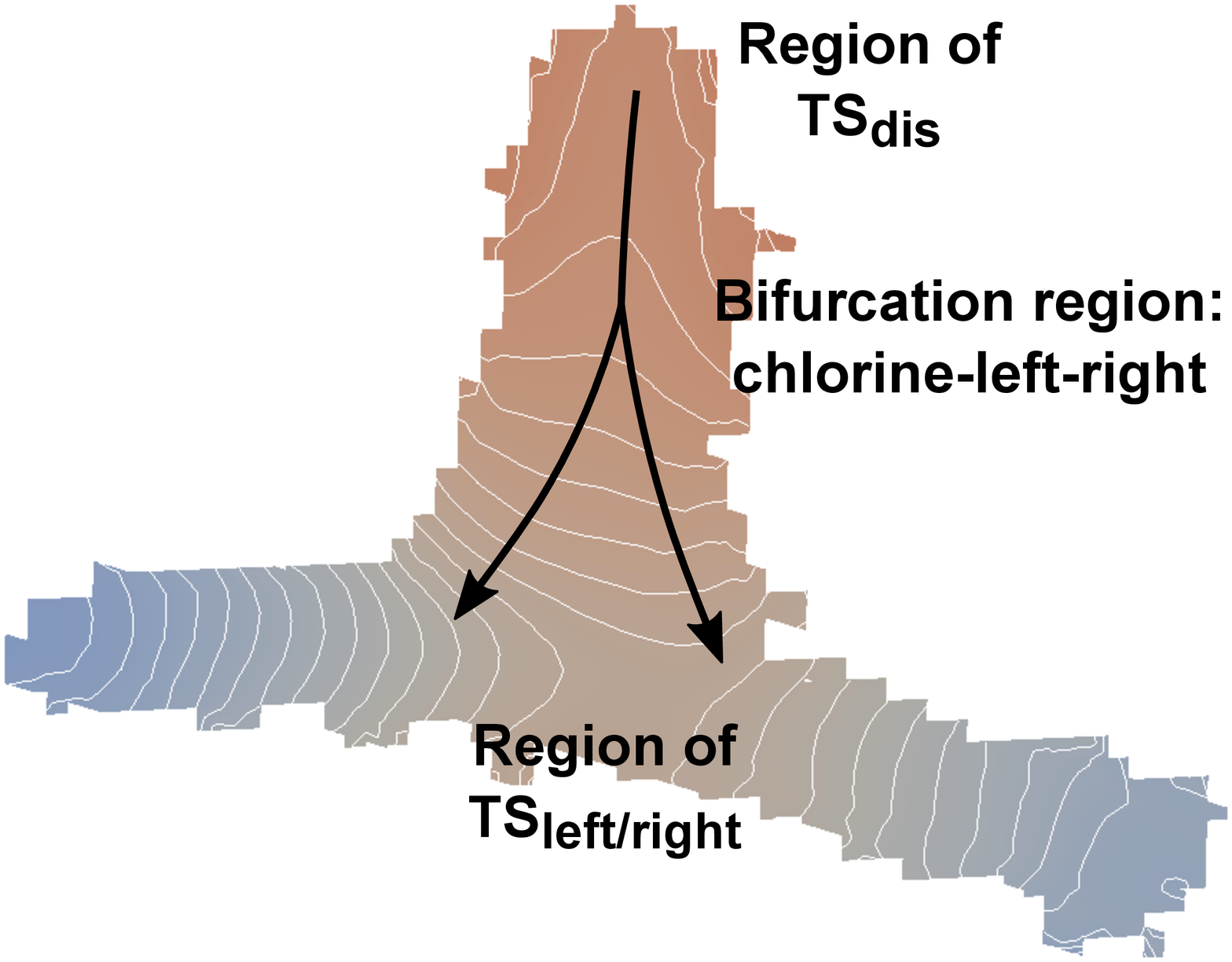}
    	\caption{$F_{0} = $ 1.0~nN}
		\label{fig:PES_DisBifurcation_1.00}
	\end{subfigure}
\hfill
	\begin{subfigure}{0.23\textwidth}
	\centering
	    \includegraphics[width=\textwidth]{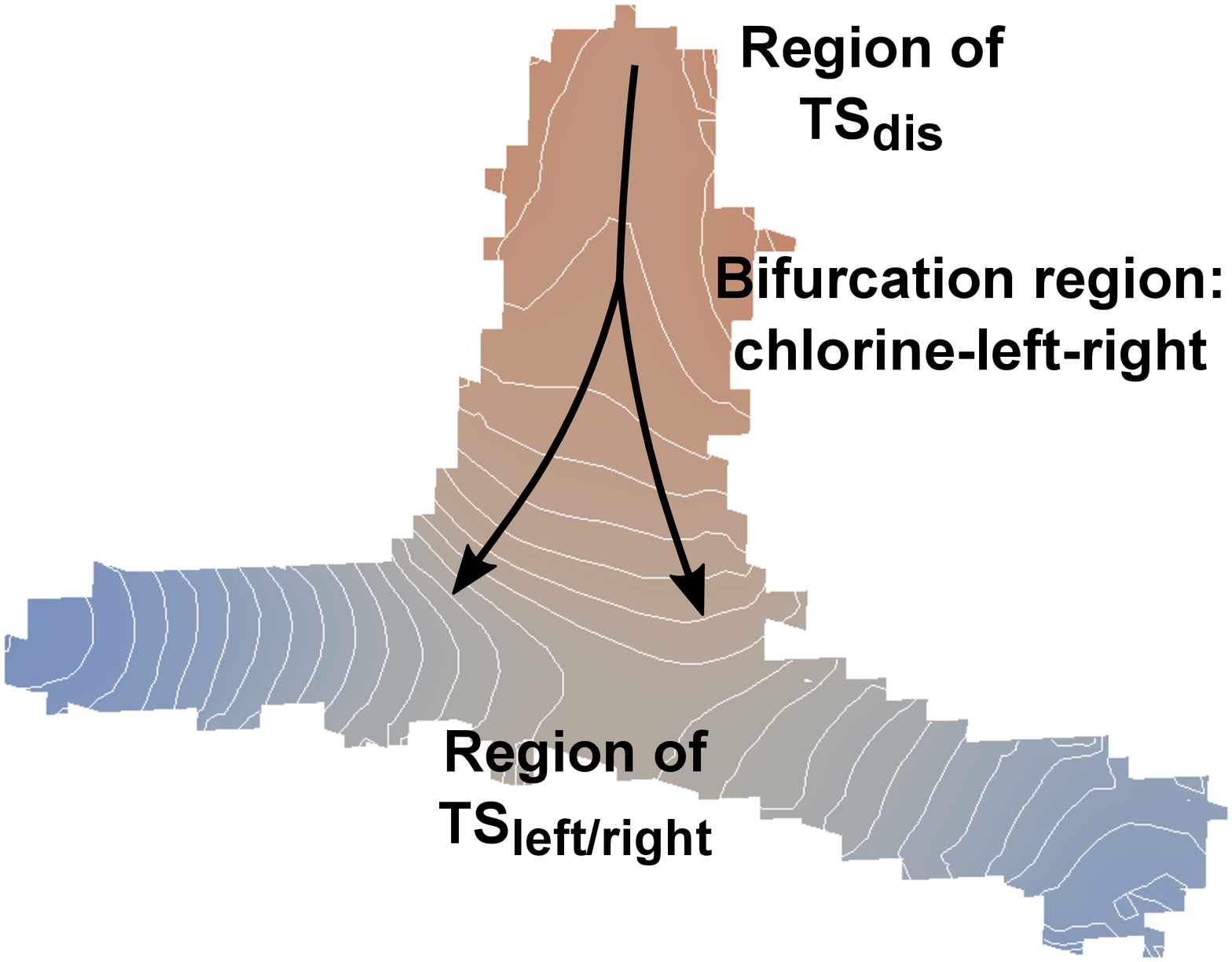}
    	\caption{$F_{0} = $ 1.5~nN}
		\label{fig:PES_DisBifurcation_1.50}
	\end{subfigure}
\hfill
	\begin{subfigure}{0.23\textwidth}
	\centering
	    \includegraphics[width=\textwidth]{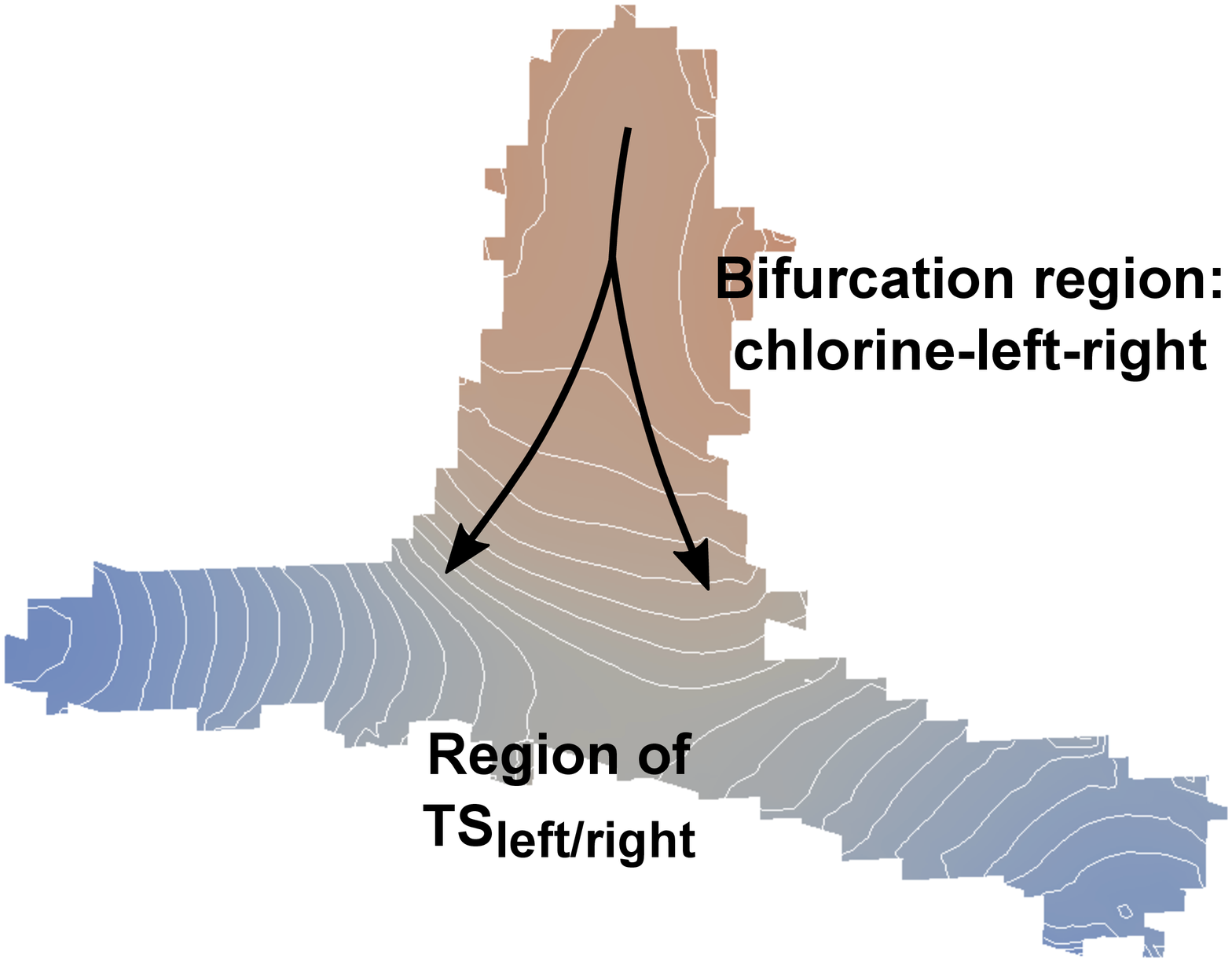}
    	\caption{$F_{0} = $ 2.0~nN}
		\label{fig:PES_DisBifurcation_2.00}
	\end{subfigure}
\hfill
	\begin{subfigure}{0.23\textwidth}
	\centering
	    \includegraphics[width=\textwidth]{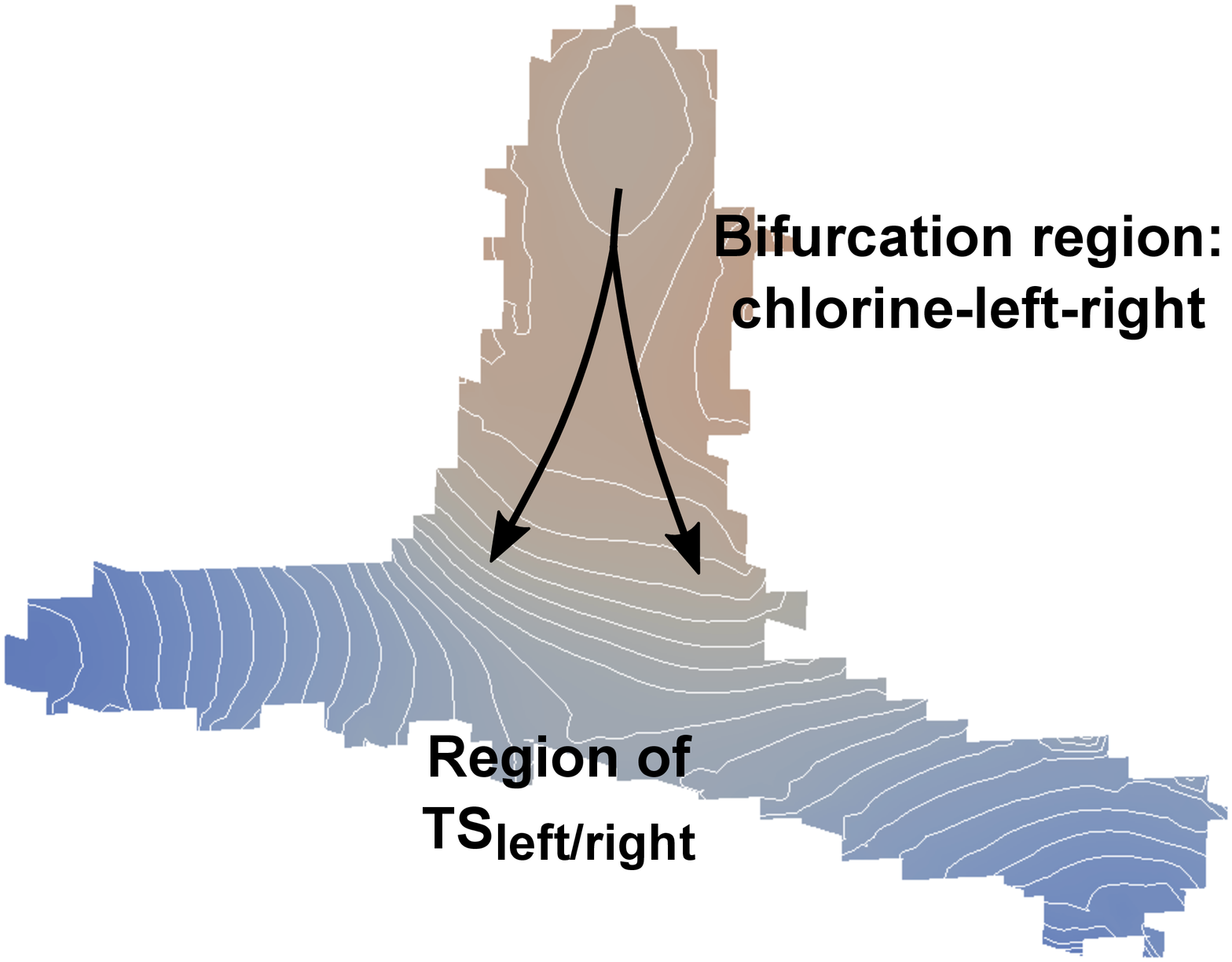}
    	\caption{$F_{0} = $ 3.0~nN}
		\label{fig:PES_DisBifurcation_3.00}
	\end{subfigure}
    \caption{Schematic illustration of the Dijkstra
paths for left/right migration of chlorine after disrotatory ring-opening 
on a set of 2D-slices of the 3D FT-ePES obtained at different constant forces
from the thermally activated limit at 0~nN in panel~(a) up to 3~nN in (e).
The $x$-axis corresponds to the Cl~position (CV3) whereas the
$y$-axis is a suitable combination of
the length of the breaking carbon-carbon bond (CV1) and the methyl dihedral (CV2).
Starting at the disrotatory transition state (top part of each panel),
the reaction pathways lead down in energy toward the
left-right downhill bifurcation (central part) 
that connect to the dis-left and dis-right final products
(left and right bottom parts, respectively). 
The reactant energy is set to \SI{0}{\kilo\calorie\per\mole} for each FT-ePES
and the depicted energy contour lines are separated by an equidistant
spacing of \SI{2}{\kilo\calorie\per\mol}
where red encodes increasing and blue decreasing energies compared to the reactant energy.
}
	\label{fig:2D-Slices-Bifurcation_Energies}
\end{figure}

After having discussed how the reaction channels originating
in the reactant state split into dis- and conrotatory ring-opening
processes below and above the catastrophe at $F_0^{\rm crit}$, 
we finally turn to the analysis of the decision about
chlorine left/right migration in terms of the FT-ePES
which takes place in the product regime of the energy landscape. 
The 2D-slices on the product side of the disrotatory reaction region in 
Fig.~\ref{fig:2D-Slices-Bifurcation_Energies} are oriented in such a way 
that the horizontal axis corresponds to the Cl~position 
that nicely indicates left/right migration of Cl7, 
whereas the vertical axis results from a suitable combination of
the distance of the breaking carbon-carbon bond and the methyl dihedral. 
In the following digression, 
we will concentrate on the bifurcation between dis-left- and dis-right-products, 
because the corresponding bifurcation on the conrotatory part of the FT-ePESs is 
essentially symmetric due to the underlying
structural symmetries along this ring-opening channel,
which are not broken by applying colinear external forces at the two methyl groups.

In the thermal reference case at 0~nN, 
the disrotatory transition state is located in the top part of the 2D-slice 
from there on the reaction paths evolve down in energy, see
Fig.~\ref{fig:2D-Slices-Bifurcation_Energies}(a). 
The reaction channel in the middle part of the shown FT-ePES is relatively flat 
in the direction perpendicular to the course of the reaction, thus
facilitating the structural separation of the two paths toward left and right migration.
The resulting dis-left- and dis-right products in the bottom left/right 
corners of the FT-ePES are separated by the second transition state, 
TS$_{\rm left/right}$ (being analogous to the generic $\textrm{TS}_{\textrm{1-2}}$ 
depicted in panel~(a) of Fig.~\ref{fig:Bifurcations_Generell}), 
that is clearly visible in the bottom-middle area in terms of the pattern
of the contour lines.

As expected, the zero-force ePES gets distorted both in energy and shape due to the
application of the tensile force to the methyl groups.
With increasing force, the bifurcation area becomes flatter, which can be easily
estimated by the decreasing number of contour lines when going from (a) to (b) etc. 
Furthermore, the position of the second transition state, which directly connects 
the dis-left to the dis-right product, shifts when applying a mechanical force.
The second transition state (region) is located to a greater extent 
close to the center of the bifurcation area, making the surface more symmetric.
Thus, in the case of a sufficiently large force and at finite temperatures, the 
force-transformed energy landscape, FT-ePES, 
would lead to a higher amount of the dis-right product
compared to the thermal situation at zero force.
Clearly, after the catastrophe on the reactant side of the energy landscape
occurred, i.e. at forces exceeding $F_{0}^{\rm{crit}}$,
the isolines in the top part of 
Fig.~\ref{fig:2D-Slices-Bifurcation_Energies}(d) and~(e) 
do no longer correspond to a transition state as they did
at the lower forces in panels~(a) to (c).

Overall, it is found that the region of that downhill bifurcation 
which decides about 
the left/right direction of the migrating chlorine atom, Cl7, does 
only change in quantitative details when applying tensile forces to the methyl 
groups, whereas the local topology of this part of the 
energy landscape remains invariant
in agreement with the two disrotatory branches in the lower
parts of panels~(d) and~(e) in Fig.~\ref{fig:ReactionSchemes}
that connect to the dis-left and dis-right products.

\section{Conclusions and Outlook}
\label{sec:Conclusions_and_outlook}

It is well appreciated in the extant literature that
dihalogenated cyclopropanes display unexpected chemistry
upon mechanochemical ring-opening reactions. 
In an effort to shed light on this puzzle, we investigated in the first step
of the present study 
the impact of halogen substitution on the thermally activated 
ring-opening process followed by halogen migration of 
(2\textit{S},3\textit{S})-1,1-dihalo-2,3-dimethylcyclopropanes 
in the case of~F, Cl, Br, and~I disubstitution. 
Surprisingly, the decision about the direction of halogen migration 
(called ``left'' and ``right'' for simplicity) 
in case of disrotatory ring-opening
occurs in qualitatively different ways for this 
difluorinated cyclopropane compared to the homologous 
Cl, Br, and I~species. 
For \textit{trans}-\textit{gem}-difluorocyclopropane, 
the left/right decision is governed by an 
uphill bifurcation followed by two separate transition states for each migration
direction, whereas the disrotatory transition state is
first surmounted in the other three cases followed by 
a downhill bifurcation into the left and right disrotatory products. 
This change in the bifurcation scenario
is corroborated by vibrational mode analysis performed along 
the respective intrinsic reaction coordinates. 
Thus, chemical substitution is found to qualitatively change
the potential energy landscape for the ring-opening of cyclopropanes. 

In our second step, which is based on quantum mechanochemical computations
at constant external forces, 
we 
discover that the topology of the energy landscape for 
mechanochemically activated ring-opening of 
\textit{trans}-\textit{gem}-dichlorocyclopropane
is qualitatively different in the high force regime compared
to low forces 
down to the thermal limit.
In particular, an uphill bifurcation that decides about
dis- {\em versus} conrotatory ring-opening at low forces
gets transmuted at sufficiently high forces
such that first, one transition state common to both dis- and conrotatory
pathways is encountered followed by a first downhill bifurcation
that decides about dis- and conrotatory ring-opening
and, subsequently, by two additional separate downhill bifurcations
for left/right migration.

The two topologically distinct low- and high-force energy landscapes 
are separated from each other by a cusp catastrophe,
which happens at a critical force $F_0^{\rm crit}$
of roughly 1.6~nN in the present case. 
Analysis of the energy profiles as a function of external force
in conjunction with topological analysis 
reveals that disrotatory ring-opening is energetically
preferred below $F_0^{\rm crit}$ down to the zero force
limit (where it becomes the 
Woodward-Hoffmann allowed electrocyclic reaction), 
whereas the conrotatory and disrotatory processes possess the same transition state
in the high force limit
beyond $F_{0}^{\rm{crit}}$.
Thus, we find here that it is an intricate topological catastrophe that 
underlies and 
%
therefore
explains the previously predicted~\cite{Wollenhaupt2015} change in
the ring-opening mechanism of 
\textit{trans}-\textit{gem}-dichlorocyclopropanes
at about 1.6~nN in accord with
subsequent 
single-molecule force spectroscopy experiments.\cite{Wang2015}

In more general terms, sequential transition states
and the branching of reaction pathways {\em via} 
bifurcations that lead to multiple product channels
are well known in the study of chemical reactivity. 
Here, we show that both uphill and downhill bifurcations
being pre- and post-transition-state features, respectively, 
can be shifted {\em continuously} by applying external
mechanical forces as typically generated by sonication techniques. 
These changes can be merely quantitative in certain
force regimes, but they can also qualitatively change
the topology of the energy landscape that describes the
process, for instance by shifting an uphill bifurcation
over transition states to the downhill side. 
Beyond the specific case, the present findings delineate an approach 
towards steering the selectivity of chemical reactions 
in favor of desired products by means of mechanochemical activation. 

\section{Methods}
\label{sec:Methods}
Optimization of both, transition states and IRCs of the four 
halogen-substituted cyclopropane derivatives 
have been performed using the unrestricted BLYP density functional
and the TZVP basis set as implemented in Gaussian~09.\cite{g09-C1_KurzForm}
For the
specific case of the
dichloro
cyclopropane derivative investigated in this study, 
good agreement
between BLYP and
CASSCF calculations can be observed as demonstrated
in the SI of Ref.~\citenum{Wollenhaupt2015}
and supported by previous benchmark calculations for bond
breaking in mechanochemical reactions.\cite{Iozzi2009}
In addition, the
NEVPT2 calculations in Sec.~1.2 of the SI of the present work
for the difluoro species support this conclusion.
A detailed description of 
the computational
methods
can be found in Sec.~1.3 of the SI.
The force-transformed potential energy landscapes (FT-PES) 
have been generated by means of
\textit{ab initio} molecular dynamics simulations~\cite{Marx2009} 
on the basis of the Car-Parrinello method.\cite{Car1985} 
The reduced-dimensionality versions (FT-ePES) have been analyzed in terms of 
reaction pathways, not only including transition states but also bifurcation points, 
with the help of the so-called Dijkstra algorithm,\cite{Dijkstra1959}
thus providing what we call Dijkstra paths. 
An extensive discussion of the respective computational details 
can be found in Sects.~1.5 and 1.6 of the SI.

\section{Acknowledgement}

It gives us pleasure to thank Przemyslaw Dopieralski 
and Michael R\"omelt for helpful discussions.
We gratefully acknowledge generous financial support by the
Reinhart Koselleck Grant
"{}Understanding Mechanochemistry{}" (DFG MA~1547/9)
as well as computer resources 
at HLRS Stuttgart, {\sc BoViLab}@RUB, and RV--NRW. 
%

%
%
%

%
%
%
%

%
\input{manuscript.bbl}

%
%
%
%

\end{document}

%% file: manuscript.bbl
%